\documentclass[journal,12pt,onecolumn]{IEEEtran}

\usepackage{setspace}
\ifCLASSOPTIONonecolumn \doublespace \fi
\usepackage{graphics}
\usepackage{rotating}
\usepackage{amsfonts}
\usepackage{amsmath}
\usepackage{subfigure}
\usepackage{multirow}
\allowdisplaybreaks[1]

\graphicspath{{Figures/}} \DeclareGraphicsExtensions{.eps,.ps}

\begin{document}

\title{Diversity Analysis of Bit-Interleaved Coded Multiple Beamforming}

\author{\IEEEauthorblockN{Hong Ju Park and Ender Ayanoglu}\\
\IEEEauthorblockA{Center for Pervasive Communications and Computing\\
Department of Electrical Engineering and Computer Science\\
University of California, Irvine\\
Email: hjpark@uci.edu, ayanoglu@uci.edu}}

\maketitle

\ifCLASSOPTIONonecolumn
 \setlength\arraycolsep{4pt}
\else
 \setlength\arraycolsep{2pt}
\fi

\begin{abstract}
In this paper, diversity analysis of bit-interleaved coded multiple
beamforming (BICMB) is extended to the case of general spatial
interleavers, removing a condition on their previously known design
criteria and quantifying the resulting diversity order. The
diversity order is determined by a parameter $Q_{max}$ which is
inherited from the convolutional code and the spatial de-multiplexer
used in BICMB. We introduce a method to find this parameter by
employing a transfer function approach as in finding the weight
spectrum of a convolutional code. By using this method, several
$Q_{max}$ values are shown and verified to be identical with the
results from a computer search. The diversity analysis and the
method to find the parameter are supported by simulation results. By
using the Singleton bound, we also show that $Q_{max}$ is lower
bounded by the product of the number of streams and the code rate of
an encoder. The design rule of the spatial de-multiplexer for a
given convolutional code is proposed to meet the condition on the
maximum achievable diversity order.
\end{abstract}

\section{Introduction} \label{sec:introduction}

When the channel information is perfectly available at the
transmitter, beamforming is an attractive technique to enhance the
performance of a multi-input multi-output (MIMO) system
\cite{jafarkhaniBook}. A set of beamforming vectors is obtained by
singular value decomposition (SVD) which is optimal in terms of
minimizing the average bit error rate (BER) \cite{palomarTSP03}.
Single beamforming, which carries only one symbol at a time, was
shown to achieve full diversity order of $NM$ where $N$ is the
number of transmit antennas and $M$ is the number of receive
antennas \cite{sengulTC06AnalSingleMultpleBeam},
\cite{OrdonezTSP07}. However, multiple beamforming, which increases
the throughput by sending multiple symbols at a time, loses the full
diversity order over flat fading channels.

To achieve the full diversity order as well as the full spatial
multiplexing order, bit-interleaved coded multiple beamforming,
combining bit-interleaved coded modulation (BICM) and multiple
beamforming, was introduced in \cite{akayTC06BICMB}. Design criteria
for interleaving the coded sequence were provided such that each
subchannel created by SVD is utilized at least once with a
corresponding channel bit equal to $1$ in an error event on the
trellis diagram \cite{akayTC06BICMB}, \cite{akayTC06BICMB_arxiv}.
BICMB with $1/2$-rate convolutional encoder, a simple interleaver
and soft-input Viterbi decoder was shown to have full diversity
order when it is used in a $2 \times 2$ system with $2$ streams. In
this paper, the diversity order is analyzed even when the
interleaver does not meet the criteria of \cite{akayTC06BICMB},
\cite{akayTC06BICMB_arxiv}. To determine the diversity order, the
error events that dominate BER performance need to be found. We
introduce a method to find the dominant error events by extending a
method from convolutional code analysis to determine system
performance, e.g., \cite{proakis}, \cite{haccounTCOM89}, into the
analysis of the given combination of the interleaver and the code.
We also show that for any convolutional code and any spatial
de-multiplexer, the maximum achievable diversity order is related
with the product of the code rate and the number of streams, by
using the Singleton bound \cite{SingletonTIT64}. The design rule of
the spatial de-multiplexer to get the maximum achievable diversity
order is also proposed.

The rest of this paper is organized as follows. A brief review of
the BICMB system is given in Section \ref{sec:system_model}. Section
\ref{sec:alpha_Spectra} introduces a method to find $\alpha$-vectors
for a given convolutional code and the number of subchannels.
Pairwise error probability (PEP) analysis is given in Section
\ref{sec:cal_order}. In Section \ref{sec:do_bound}, the analysis of
the maximum achievable diversity order of BICMB is shown, and the
design rule of the spatial de-multiplexer for the maximum achievable
diversity order is proposed. Simulation results supporting the
analysis are shown in Section \ref{sec:result}. Finally, we end the
paper with a conclusion in \ref{sec:conclusion}.

\section{BICMB Overview} \label{sec:system_model}

The code rate $R_c = k_c / n_c$ convolutional encoder, possibly
combined with a perforation matrix for a high rate punctured code,
generates the codeword $\mathbf{c}$ from the information vector
$\mathbf{b}$. Then, the spatial de-multiplexer distributes the coded
bits into $S$ sequences, each of which is interleaved by an
independent bit-wise interleaver. The interleaved sequences
$\mathbf{D}$ are mapped by Gray encoding onto the symbol sequences
$\mathbf{Y}$. A symbol belongs to a signal set $\chi \subset
\mathbb{C}$ of size $|\chi| = 2^m$, such as $2^m$-QAM, where $m$ is
the number of input bits to the Gray encoder.

The MIMO channel $\mathbf{H} \in \mathbb{C}^{M \times N}$ is assumed
to be quasi-static, Rayleigh, and flat fading, and perfectly known
to both the transmitter and the receiver. In this channel model, we
assume that the channel coefficients remain constant for the $L$
symbol duration. The beamforming vectors are determined by the
singular value decomposition of the MIMO channel, i.e., $\mathbf{H}
= \mathbf{U \Lambda V^H}$ where $\mathbf{U}$ and $\mathbf{V}$ are
unitary matrices, and $\mathbf{\Lambda}$ is a diagonal matrix whose
$s^{th}$ diagonal element, $\lambda_s \in \mathbb{R}$, is a singular
value of $\mathbf{H}$ with decreasing order. When $S$ symbols are
transmitted at the same time, then the first $S$ vectors of
$\mathbf{U}$ and $\mathbf{V}$ are chosen to be used as beamforming
matrices at the receiver and the transmitter, respectively. Let's
denote the first $S$ vectors of $\mathbf{U}$ and $\mathbf{V}$ as
$\mathbf{\tilde{U}}$ and $\mathbf{\tilde{V}}$. The system
input-output relation at the $k^{th}$ time instant for a packet
duration is written as
\begin{align}
\mathbf{r}_k = \mathbf{\tilde{U}}^H \mathbf{H} \mathbf{\tilde{V}}
\mathbf{y}_k + \mathbf{\tilde{U}}^H \mathbf{n}_k
\label{eq:System_InOut}
\end{align}
where $\mathbf{y}_k$ is an $S \times 1$ vector of transmitted
symbols, $\mathbf{r}_k$ is an $S \times 1$ vector of the detected
symbols, and $\mathbf{n}_k$ is an additive white Gaussian noise
vector with zero mean and variance $N_0 = N / SNR$. On each $s^{th}$
subchannel, finally, we get
\begin{align}
r_{k,s} = \lambda_s y_{k,s} + \tilde{n}_{k,s}
\label{eq:Subchannel_InOut}
\end{align}
where $r_{k,s}$, $y_{k,s}$, and $\tilde{n}_{k,s}$ are a detected
symbol, a transmitted symbol, and a noise term, respectively.
$\mathbf{H}$ is complex Gaussian with zero mean and unit variance,
and to make the received signal-to-noise ratio $SNR$, the total
transmitted power is scaled as $N$. The equivalent system model is
shown in Fig. \ref{fig:eq_bicmb}.

\ifCLASSOPTIONonecolumn
\begin{figure}[!m]
\centering \includegraphics[width = 0.75\linewidth]{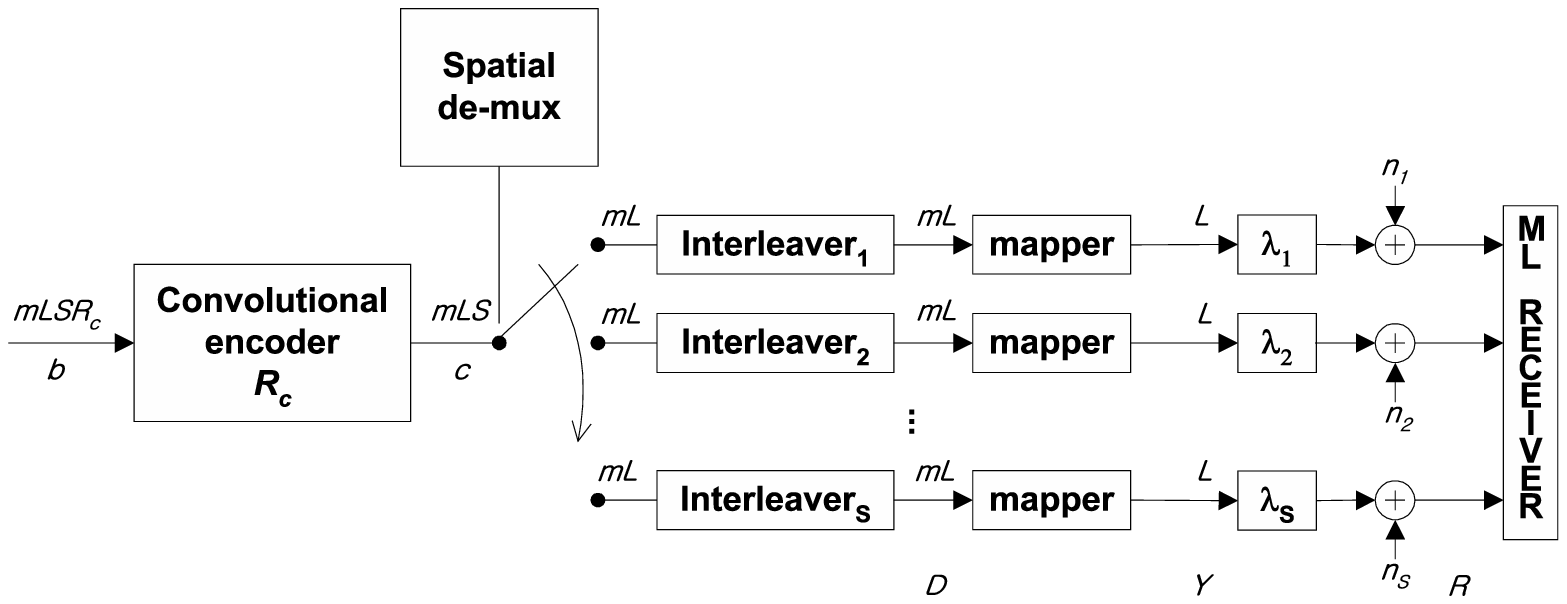}
\caption{Equivalent system model of BICMB} \label{fig:eq_bicmb}
\end{figure}
\else
\begin{figure}[!t]
\centering \includegraphics[width = 1\linewidth]{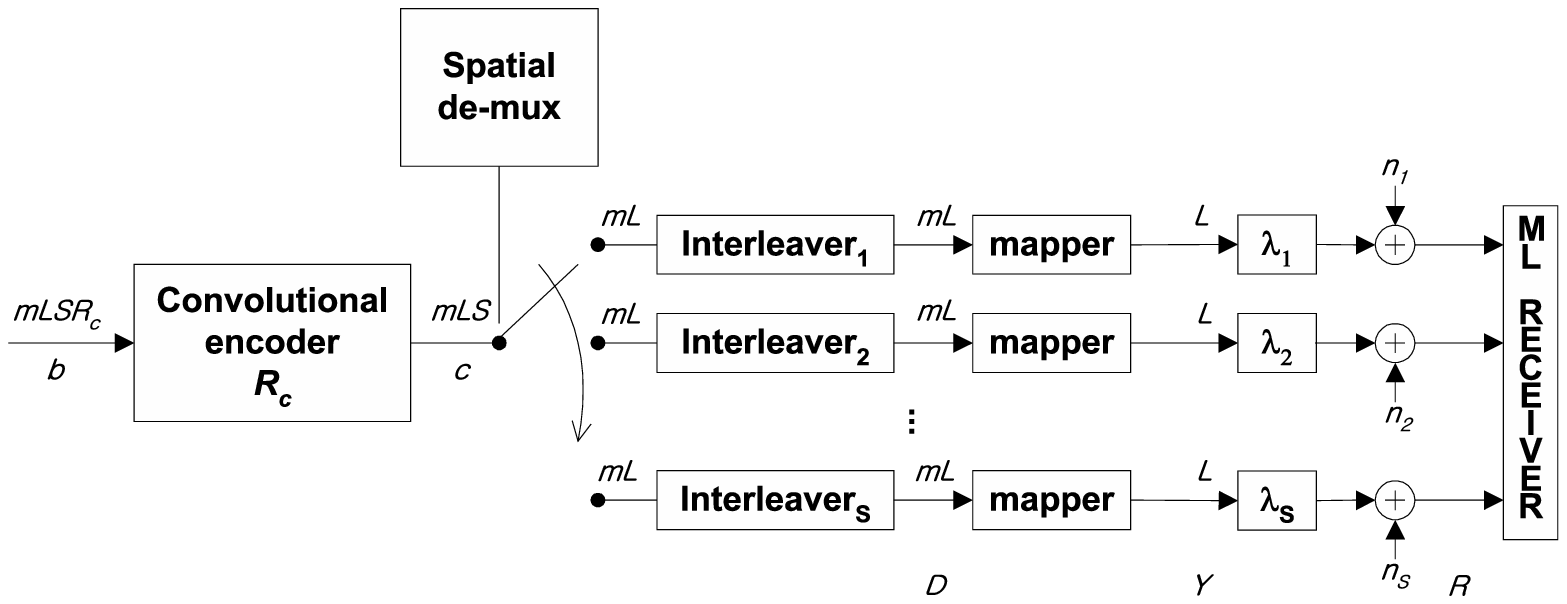}
\caption{Equivalent system model of BICMB} \label{fig:eq_bicmb}
\end{figure}
\fi

The location of the $l^{th}$ coded bit $c_l$ within the detected
symbols is stored in a table $l \rightarrow (k, s, i)$, where $k$,
$s$, and $i$ are time instant, subchannel, and bit position on a
symbol, respectively. Let $\chi_{b}^{i} \subset \chi$ where $b \in
\{0, 1\}$ in the $i^{th}$ bit position. By using the information in
the table and the input-output relation in
(\ref{eq:Subchannel_InOut}), the receiver calculates the ML bit
metrics as
\begin{align}
\gamma^i(r_{k,s}, c_l) = \min_{y \in \chi_{c_l}^{i}} |r_{k,s} -
\lambda_s y |^2. \label{eq:ML_bit_metrics}
\end{align}
The combination of the ML bit metrics of (\ref{eq:ML_bit_metrics})
and $\mathbf{\tilde{U}}$ detector at the receiver is not the unique
solution to get the optimum BER performance. Appropriate bit metrics
corresponding to a linear detector, such as zero-forcing (ZF) or
minimum mean square error (MMSE) detector, were shown to be
equivalent to the bit metrics of (\ref{eq:ML_bit_metrics}) with
$\mathbf{\tilde{U}}$ detector \cite{ErsinTCOMM08}. Finally, the ML
decoder can make decisions according to the rule
\begin{align}
\mathbf{\hat{c}} = \arg\min_{\mathbf{\tilde{c}}} \sum_l
\gamma^i(r_{k,s}, \tilde{c}_l). \label{eq:Decision_Rule}
\end{align}

\section{$\alpha$-Spectra} \label{sec:alpha_Spectra}

The BER of a BICMB system is upper bounded by all the summations of
each pairwise error probability for all the error events on the
trellis \cite{akayTC06BICMB}, \cite{akayTC06BICMB_arxiv}. Therefore,
the calculation of PEP for each error event is needed to analyze the
diversity order of a given BICMB system. If the interleaver is
properly designed such that the consecutive long coded bits are
mapped onto distinct symbols, the PEP between the two codewords
$\mathbf{c}$ and $\mathbf{\hat{c}}$ with Hamming distance $d_H$ is
upper bounded as \cite{akayTC06BICMB}
\begin{align}
P\left(\mathbf{c} \rightarrow \mathbf{\hat{c}}\right) &=
E\left[P\left(\mathbf{c} \rightarrow \mathbf{\hat{c}}|
\mathbf{H} \right)\right] \nonumber \\
& \leq E \left[ \frac{1}{2} \exp \left(- \frac{d^2_{min}
\sum\limits_{s=1}^{S} \alpha_{s} \lambda_{s}^2}{4 N_0} \right)
\right] \label{eq:PEP_given_H}
\end{align}
where $d_{min}$ is the minimum Euclidean distance in the
constellation and $\alpha_s$ denotes the number of times the
$s^{th}$ subchannel is used corresponding to $d_H$ bits under
consideration, satisfying $\sum_{s=1}^{S} \alpha_s = d_H$. Since PEP
is affected by the summation of the products between $\alpha_s$ and
singular values as can be seen in (\ref{eq:PEP_given_H}), it is
important to calculate the $\alpha$-vectors for each error path to
have an insight into the diversity order behavior of a particular
BICMB implementation.

It has been shown in \cite{akayTC06BICMB},
\cite{akayTC06BICMB_arxiv} that for a single-carrier BICMB system,
if the interleaver is designed such that, for all error paths of
interest with Hamming distance $d_H$ to the all-zeros path,
\begin{enumerate}
\item the consecutive coded bits are mapped over different symbols,
\item $\alpha_s \geq 1$ for $1 \leq s \leq S$,
\end{enumerate}
then the BICMB system achieves full diversity. In this paper, we
will analyze cases where the sufficient condition $\alpha_s \geq 1$
may not be satisfied, i.e., $\alpha_s = 0$ for some $s=1,2,\cdots,S$
is possible. In order to carry out this analysis, as well as to get
an insight into the system behavior in \cite{akayTC06BICMB},
\cite{akayTC06BICMB_arxiv}, one needs a method to calculate the
values of $\alpha_s$ (which we call the $\alpha$-vector) of an error
path at Hamming distance $d_H$ to the all-zeros path.

We developed a method to calculate the $\alpha$-vectors for a
convolutional code and interleaver combination. We will now
illustrate this method with a simple example. For this example, the
system is composed of a $4$-state $1/2$-rate convolutional encoder
and a spatial de-multiplexer rotating with an order of $a$, $b$,
$c$, and $d$ which represent the four streams of transmission. Fig.
\ref{fig:trellis_alpha} represents a trellis diagram of this
convolutional encoder for one period at the steady state. Since a
convolutional code is linear, the all-zero codeword is assumed to be
the input to the encoder. To find a transfer function of a
convolutional code and a spatial de-multiplexer, we label the
branches as a combination of $a^{\phi_a}$, $b^{\phi_b}$,
$c^{\phi_c}$, and $d^{\phi_d}$, where the exponent $\phi_i$ denotes
the number of usage of the subchannel $i$ which contributes to
detecting the wrong branch by the detector. Additionally,
$Z^{\phi_Z}$, whose exponent satisfies $\phi_Z =
\phi_a+\phi_b+\phi_c+\phi_d$, is included to get the relationship
between the Hamming distance $d_H$ and $\alpha$-vector of an error
event. Furthermore, the non-zero states are arbitrarily labeled
$X_{11}$ through $X_{23}$, while the zero state is labeled as $X_i$
if branches split and $X_o$ if branches merge as shown in Fig.
\ref{fig:trellis_alpha}.

\ifCLASSOPTIONonecolumn
\begin{figure}[!m]
\centering \includegraphics[width =
0.3\linewidth]{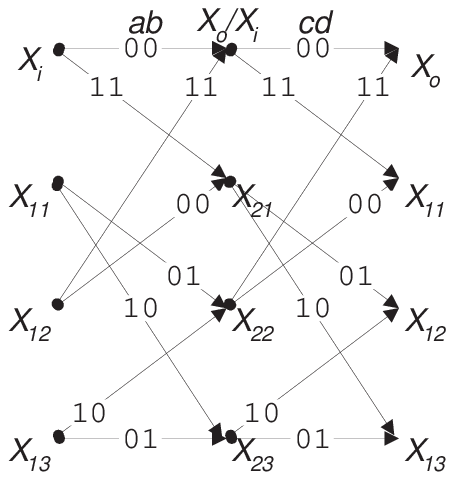} \caption{Trellis of $4$-state
$1/2$-rate convolutional code with $4$ streams}
\label{fig:trellis_alpha}
\end{figure}
\else
\begin{figure}[!t]
\centering \includegraphics[width =
0.6\linewidth]{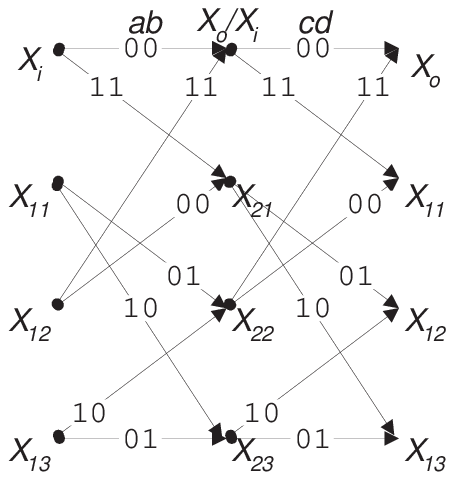} \caption{Trellis of $4$-state
$1/2$-rate convolutional code with $4$ streams}
\label{fig:trellis_alpha}
\end{figure}
\fi

Let's denote $\mathbf{x}$ = $ \left[
\begin{array}{cccccc} X_{11} & X_{12} & X_{13} & X_{21} & X_{22} &
X_{23} \end{array} \right]^T$. Then, the state equations are given
by the matrix equation
\ifCLASSOPTIONonecolumn
\begin{align}
\mathbf{x} &= \mathbf{Fx} + \mathbf{t}X_i = \left[
\begin{array}{cccccc}
0 & 0 & 0 & 0 & 1 & 0 \\
0 & 0 & 0 & dZ & 0 & cZ \\
0 & 0 & 0 & cZ & 0 & dZ \\
0 & 1 & 0 & 0 & 0 & 0 \\
bZ & 0 & aZ & 0 & 0 & 0 \\
aZ & 0 & bZ & 0 & 0 & 0 \end{array} \right] \mathbf{x} + \left[
\begin{array}{c}
cdZ^2 \\ 0 \\ 0 \\ abZ^2 \\ 0 \\ 0\end{array} \right] X_{i}.
\label{eq:TransMat1}
\end{align}
\else
\begin{align}
\mathbf{x} &= \mathbf{Fx} + \mathbf{t}X_i \nonumber \\
&= \left[
\begin{array}{cccccc}
0 & 0 & 0 & 0 & 1 & 0 \\
0 & 0 & 0 & dZ & 0 & cZ \\
0 & 0 & 0 & cZ & 0 & dZ \\
0 & 1 & 0 & 0 & 0 & 0 \\
bZ & 0 & aZ & 0 & 0 & 0 \\
aZ & 0 & bZ & 0 & 0 & 0 \end{array} \right] \mathbf{x} + \left[
\begin{array}{c}
cdZ^2 \\ 0 \\ 0 \\ abZ^2 \\ 0 \\ 0\end{array} \right] X_{i}.
\label{eq:TransMat1}
\end{align}
\fi
We also get
\begin{align}
X_{o} = \mathbf{gx} = \left[ \begin{array}{cccccc} 0 & abZ^2 & 0 & 0
& cdZ^2 & 0 \end{array} \right] \mathbf{x}. \label{eq:TransMat2}
\end{align}
The transfer function is represented in closed form by using the
method in \cite{haccounTCOM89} as
\ifCLASSOPTIONonecolumn
\begin{align} \label{eq:Transfunc_S4R1_2}
\mathbf{T}(a,b,c,d,Z) &= \mathbf{g}
\left[\mathbf{I}-\mathbf{F}\right]^{-1} \mathbf{t} =\mathbf{gt} +
\sum\limits_{k=1}^{\infty} \mathbf{gF}^{k} \mathbf{t} \nonumber\\
&= Z^5 (a^2 b^2 d + b c^2 d^2) \nonumber\\
&+ Z^6 (2 a^2 b c^2 d + a^2 b^2 d^2 + b^2 c^2 d^2) \nonumber\\
&+ Z^7 (a^2 b^3 c^2 + 2 a^2 b^2 c^2 d + 2 a^2 b c^2 d^2 + \\
&\quad\qquad b^3 c^2 d^2 + a^2 b^2 d^3 + a^2 c^2 d^3) \nonumber\\
&+ Z^8 (a^4 b^2 c^2 + 4 a^2 b^3 c^2 d + 4 a^2 b^2 c^2 d^2 + \nonumber\\
&\quad\qquad b^4 c^2 d^2 + a^2 c^4 d^2 + 4 a^2 b c^2 d^3 + a^2 b^2
d^4) + \cdots \nonumber
\end{align}
\else
\begin{align} \label{eq:Transfunc_S4R1_2}
\mathbf{T}&(a,b,c,d,Z) = \mathbf{g}
\left[\mathbf{I}-\mathbf{F}\right]^{-1} \mathbf{t} =\mathbf{gt} +
\sum\limits_{k=1}^{\infty} \mathbf{gF}^{k} \mathbf{t} \nonumber\\
&\quad= Z^5 (a^2 b^2 d + b c^2 d^2) \nonumber\\
&\quad+ Z^6 (2 a^2 b c^2 d + a^2 b^2 d^2 + b^2 c^2 d^2) \nonumber\\
&\quad+ Z^7 (a^2 b^3 c^2 + 2 a^2 b^2 c^2 d + 2 a^2 b c^2 d^2 + \\
&\qquad\qquad b^3 c^2 d^2 + a^2 b^2 d^3 + a^2 c^2 d^3) \nonumber\\
&\quad+ Z^8 (a^4 b^2 c^2 + 4 a^2 b^3 c^2 d + 4 a^2 b^2 c^2 d^2 + \nonumber\\
&\qquad\qquad b^4 c^2 d^2 + a^2 c^4 d^2 + 4 a^2 b c^2 d^3 + a^2 b^2
d^4) + \cdots \nonumber
\end{align}
\fi
where $[\mathbf{I}-\mathbf{F}]^{-1}$ can be expanded as
$\mathbf{I} + \mathbf{F} + \mathbf{F}^2 + \cdots$ through an
infinite series of power of matrices. The weight spectrum, used for
error performance analysis of convolutional codes, can be easily
determined by $\mathbf{T}(a,b,c,d,Z)\mid_{a=b=c=d=1}$ and can be
compared with the literature \cite{conanTCOM84}, \cite{beginTCOM90}.

Assume that $a$, $b$, $c$, $d$ are assigned to be the stream numbers
$1$, $2$, $3$, and $4$, respectively. We can then figure out from
the transfer function that the $\alpha$-vectors of two error events
with Hamming distance equal to $5$ are $[2\, 2\, 0\, 1]$ and $[0\,
1\, 2\, 2]$. Besides, the vectors with $\alpha_1$ equal to $0$ are
easily found by choosing the terms composed of only $b$, $c$, and
$d$, which are $[0\, 1\, 2\, 2]$, $[0\, 2\, 2\, 2]$, $[0\, 3\, 2\,
2]$, and $[0\, 4\, 2\, 2]$. No vector is found which has $\alpha_1$
= $\alpha_2$ = 0 or $\alpha_1$ = $\alpha_2$ = $\alpha_3$ = 0.

This method can be applied to any $\mathcal{K}$-state $k_c /
n_c$-rate convolutional code and $S$-stream BICMB system. If the
spatial de-multiplexer is not a random switch for the whole packet,
the period of the spatial de-multiplexer is an integer multiple of
the least common multiple (LCM) of $n_c$ and $S$. Note that we
restrict a period of the interleaver to correspond to an integer
multiple of trellis sections. Let's denote $P$ = $LCM(n_c, S)$ which
means the number of coded bits for a minimum period. Then, the
dimension of the vector $\mathbf{x}$ is $nP(\mathcal{K}-1)k_c/n_c$
where $n$ is the integer multiple for a period of interest.

By using this method, transfer functions of a $4$-state $1/2$-rate
convolutional code with generator polynomials $(5, 7)$ in octal
combined with several different de-multiplexers are shown in
(\ref{eq:Transfunc_S2R1_2}), (\ref{eq:Transfunc_S3R1_2}), and
(\ref{eq:Transfunc_S3R1_2_diffdemux}). The spatial de-multiplexer
used in $\mathbf{T}_1$ and $\mathbf{T}_2$ is a simple rotating
switch on $2$ and $3$ subchannels, respectively. For $\mathbf{T}_3$,
$i^{th}$ coded bit is de-multiplexed into subchannel
$s_{\mathrm{mod}(i, 18)+1}$ where $s_1$ = $\cdots$ = $s_6$ = $1$,
$s_7$ = $\cdots$ = $s_{12}$ = $2$, $s_{13}$ = $\cdots$ = $s_{18}$ =
$3$ and $\mathrm{mod}$ is the modulo operation. Throughout the
transfer functions, the variables $a$, $b$, and $c$ represent
$1^{st}$, $2^{nd}$, and $3^{rd}$ subchannel, respectively, in a
decreasing order of singular values from the channel matrix.
\begin{align}\label{eq:Transfunc_S2R1_2}
\mathbf{T}_1 &= Z^5 (a^2 b^3) + Z^6 (a^4 b^2 + a^2 b^4)\nonumber\\
&+ Z^7 (3 a^4 b^3 + a^2 b^5) \nonumber\\
&+ Z^8 (a^6 b^2 + 6 a^4 b^4 + a^2 b^6) \\
&+ Z^9 (5 a^6 b^3 + 10 a^4 b^5 + a^2 b^7 ) \nonumber\\
&+ Z^{10} (a^8 b^2 + 15 a^6 b^4 + 15 a^4 b^6 + a^2 b^8) + \cdots
\nonumber
\end{align}
\begin{align}\label{eq:Transfunc_S3R1_2}
\mathbf{T}_2 &= Z^5(a^2 b^2 c + a^2 b c^2 + a b^2 c^2) \nonumber\\
&+ Z^6(a^3 b^2 c + a^2 b^3 c + a^3 b c^2 + \nonumber\\
&\quad\qquad a b^3 c^2 + a^2 b c^3 + a b^2 c^3) \nonumber\\
&+ Z^7(2 a^3 b^3 c + 2 a^3 b^2 c^2 + 2 a^2 b^3 c^2 + \nonumber\\
&\quad\qquad 2 a^3 b c^3 + 2 a^2 b^2 c^3 + 2 a b^3 c^3)\\
&+ Z^8(a^5 b^3 + a^4 b^3 c + a^3 b^4 c + 2 a^4 b^2 c^2 +\nonumber\\
&\quad\qquad 3 a^3 b^3 c^2 + 2 a^2 b^4 c^2 + a^4 b c^3 + 3 a^3 b^2 c^3 +\nonumber\\
&\quad\qquad 3 a^2 b^3 c^3 + a b^4 c^3 + b^5 c^3 + a^3 b c^4 + \nonumber\\
&\quad\qquad 2 a^2 b^2 c^4 + a b^3 c^4 + a^3 c^5) + \cdots \nonumber
\end{align}
\begin{align}\label{eq:Transfunc_S3R1_2_diffdemux}
\mathbf{T}_3 &=Z^5 (a^5 + a^3 b^2 + a^2 b^3 +\nonumber\\
&\quad\qquad b^5 + a^3 c^2 + b^3 c^2 + a^2 c^3 + b^2 c^3 + c^5)\nonumber\\
&+ Z^6(a^4 b^2 + 3 a^3 b^3 + a^2 b^4 + a^4 c^2 + 3 a^2 b^2 c^2 +\nonumber\\
&\quad\qquad b^4 c^2 + 3 a^3 c^3 + 3 b^3 c^3 + a^2 c^4 + b^2 c^4) \\
&+ Z^7(2 a^4 b^3 + 2 a^3 b^4 + a^3 b^3 c + 7 a^3 b^2 c^2 +\nonumber\\
&\quad\qquad 7 a^2 b^3 c^2 + 2 a^4 c^3 + a^3 b c^3 + 7 a^2 b^2 c^3 +\nonumber\\
&\quad\qquad a b^3 c^3 + 2 b^4 c^3 + 2 a^3 c^4 + 2 b^3 c^4) + \cdots
\nonumber
\end{align}
$\mathbf{T}_1$ shows no term that lacks any of variables $a$ and
$b$, which means the interleaver satisfies the full diversity order
criterion, $\alpha_s \geq 1$ for $s$ = $1$, $2$
\cite{akayTC06BICMB}, \cite{akayTC06BICMB_arxiv}. Most of the terms
in $\mathbf{T}_2$ are comprised of three variables, $a$, $b$, and
$c$. However, three error events with Hamming distance of $8$ lack
one variable, resulting in the $\alpha$-vectors as $[5\, 3\, 0]$,
$[0\, 5\, 3]$, and $[3\, 0\, 5]$. In $\mathbf{T}_3$, many terms
missing one or two variables are observed. Especially, vectors with
$\alpha_s = 0$ for two subchannels can be found as $[5\, 0\, 0]$,
$[0\, 5\, 0]$, and $[0\, 0\, 5]$. In Section \ref{sec:cal_order}, we
present how these vectors affect the diversity order of BICMB.

\section{Diversity Analysis} \label{sec:cal_order}

Through the transfer functions in Section \ref{sec:alpha_Spectra},
we have seen interleavers which do not guarantee the full diversity
criteria. As stated previously, contrary to the assumption in
\cite{akayTC06BICMB} that $\alpha_s \geq 1$ for $s=1,2,\cdots,S$, we
assume in this paper that it is possible to have $\alpha_s = 0$ for
some $s=1,2,\cdots,S$. Let's define $\alpha_{nzmin}$ as the minimum
$\alpha$ among the nonzero $\alpha$'s in the $\alpha$-vector. Using
the inequality $\sum_{s=1}^{S} \alpha_{s} \lambda_{s}^2 \geq
\alpha_{nzmin} \sum_{k=1,\alpha_k \neq 0}^{S} \lambda_{k}^2$, PEP in
(\ref{eq:PEP_given_H}) can be expressed as
\begin{align}
P\left(\mathbf{c} \rightarrow \mathbf{\hat{c}}\right) \leq E \left[
\frac{1}{2} \exp \left(- W \sum\limits_{k=1}^{K} \mu_{\ell{(k)}}
\right) \right] \label{eq:PEP_nzmin}
\end{align}
where $W = d^2_{min} \alpha_{nzmin}/(4 N_0)$, $K$ is the number of
nonzero $\alpha$'s, $\ell(k)$ is an index to indicate the $k^{th}$
nonzero $\alpha$, and $\mu_s = \lambda_s^2$. To solve
(\ref{eq:PEP_nzmin}), we need the marginal pdf $f \left(
\mu_{\ell(1)}, \cdots, \mu_{\ell(K)} \right )$ by calculating
\begin{multline}
f \left( \mu_{\ell(1)}, \cdots, \mu_{\ell(K)} \right ) =
\ifCLASSOPTIONtwocolumn \\ \fi
\int_0^{\infty} \int_0^{\mu_1} \cdots
\int_0^{\mu_{\ell(1)-2}} \int_0^{\mu_{\ell(1)}} \cdots
\int_0^{\mu_{N-1}}
\rho \left(\mu_1, \cdots, \mu_N \right) \\
\times d_{\mu_N} \cdots d_{\mu_{\ell(1)+1}} d_{\mu_{\ell(1)-1}}
\cdots d_{\mu_2} d_{\mu_1}. \label{eq:Marginal_PDF}
\end{multline}
The joint pdf $\rho \left(\mu_1, \cdots, \mu_N \right)$ in
(\ref{eq:Marginal_PDF}) is available in the literature
\cite{edelmanThesis}, \cite{Verdubook} as
\begin{align}
\rho \left(\mu_1, \cdots, \mu_N \right) = p\left(\mu_1, \cdots,
\mu_N \right) e^{-\sum\limits_{i=1}^N \mu_i} \label{eq:Joint_PDF}
\end{align}
where the polynomial $p\left(\mu_1, \cdots, \mu_N \right)$ is
\begin{align}
p\left(\mu_1, \cdots, \mu_N \right) = \prod^N_{i=1} \mu_{i}^{M-N}
\prod^N_{j>i} \left (\mu_i - \mu_j \right )^2.
\label{eq:Polynomial_p}
\end{align}
Because we are interested in the exponent of $W$, the constant,
which appears in the literature, is ignored in
(\ref{eq:Polynomial_p}) for brevity.

Let's introduce \mbox{$\hat{f} \left( \mu_{\ell(1)}, \cdots,
\mu_{\ell(K)} \right )$} which is defined as
\begin{multline}
\hat{f} \left( \mu_{\ell(1)}, \cdots, \mu_{\ell(K)} \right ) =
\ifCLASSOPTIONtwocolumn \\ \fi
\int_0^{\infty} \int_0^{\mu_1} \cdots
\int_0^{\mu_{\ell(1)-2}} \int_0^{\mu_{\ell(1)}} \cdots
\int_0^{\mu_{N-1}}
\hat{\rho} \left(\mu_1, \cdots, \mu_N \right) \\
\times d_{\mu_N} \cdots d_{\mu_{\ell(1)+1}} d_{\mu_{\ell(1)-1}}
\cdots d_{\mu_2} d_{\mu_1} \label{eq:Marginal_PDF_2}
\end{multline}
where $\hat{\rho} \left( \mu_1, \cdots, \mu_N \right)$ is defined as
\ifCLASSOPTIONonecolumn
\begin{align}
\hat{\rho} \left(\mu_1, \cdots, \mu_N \right) =
\left\{
\begin{array}{ll}
p\left(\mu_1, \cdots, \mu_N \right) e^{-\left(\mu_1 +
\sum\limits_{i=1}^K \mu_{\ell(i)}\right)}
& \textrm{if $\alpha_1 = 0$} \\
p\left(\mu_1, \cdots, \mu_N \right) e^{-\sum\limits_{i=1}^K
\mu_{\ell(i)}} & \textrm{if $\alpha_1 > 0$}.
\end{array}
\right. \label{eq:Joint_PDF_2}
\end{align}
\else
\begin{multline}
\hat{\rho} \left(\mu_1, \cdots, \mu_N \right) = \\
\left\{
\begin{array}{ll}
p\left(\mu_1, \cdots, \mu_N \right) e^{-\left(\mu_1 + \sum\limits_{i=1}^K \mu_{\ell(i)}\right)} \\
& \textrm{if $\alpha_1 = 0$} \\
p\left(\mu_1, \cdots, \mu_N \right) e^{-\sum\limits_{i=1}^K
\mu_{\ell(i)}} & \textrm{if $\alpha_1 > 0$}.
\end{array}
\right. \label{eq:Joint_PDF_2}
\end{multline}
\fi Then, we can see that $\rho \left(\mu_1, \cdots, \mu_N \right)
\leq \hat{\rho} \left(\mu_1, \cdots, \mu_N \right)$ for either case
of $\alpha_1=0$ or $\alpha_1 > 0$ because \mbox{$e^{-\mu_i} \leq 1$}
for any $i$, and therefore \mbox{$f \left( \mu_{\ell(1)}, \cdots,
\mu_{\ell(K)} \right ) \leq \hat{f} \left( \mu_{\ell(1)}, \cdots,
\mu_{\ell(K)} \right )$}. The expressions for $\hat{\rho}
\left(\mu_1, \cdots, \mu_N \right)$ in (\ref{eq:Joint_PDF_2})
provide a convenience that is useful for the integration in
(\ref{eq:Marginal_PDF_2}) by removing the exponential factors
irrelevant to the variables of integration.

For any case of $\hat{\rho} \left(\mu_1, \cdots, \mu_N \right)$ in
(\ref{eq:Joint_PDF_2}), $\hat{f}\left( \mu_{\ell(1)}, \cdots,
\mu_{\ell(K)} \right )$ can be decomposed into two polynomials as
\ifCLASSOPTIONonecolumn
\begin{align}
\hat{f} \left( \mu_{\ell(1)}, \cdots, \mu_{\ell(K)} \right ) = h
\left(\mu_{\ell(1)}, \cdots, \mu_{\ell(K)} \right ) \times g\left(
\mu_{\ell(1)}, \cdots, \mu_{\ell(K)} \right )
e^{-\sum\limits_{k=1}^K \mu_{\ell(k)}}.
\label{eq:Joint_PDF_decomposed}
\end{align}
\else
\begin{multline}
\hat{f} \left( \mu_{\ell(1)}, \cdots, \mu_{\ell(K)} \right ) = \\
h \left(\mu_{\ell(1)}, \cdots, \mu_{\ell(K)} \right ) \times g\left(
\mu_{\ell(1)}, \cdots, \mu_{\ell(K)} \right )
e^{-\sum\limits_{k=1}^K \mu_{\ell(k)}}.
\label{eq:Joint_PDF_decomposed}
\end{multline}
\fi
The polynomial $g \left( \mu_{\ell(1)}, \cdots, \mu_{\ell(K)}
\right)$ consists of factors irrelevant to the integration as
\begin{align}
g \left( \mu_{\ell(1)}, \cdots, \mu_{\ell(K)} \right) =
\prod^K_{k=1} \mu_{\ell(k)}^{M-N} \prod^K_{j>k} \left( \mu_{\ell(k)}
- \mu_{\ell(j)} \right )^2. \label{eq:function_g}
\end{align}
The other polynomial $h \left( \mu_{\ell(1)}, \cdots, \mu_{\ell(K)}
\right)$ for $\alpha_1 = 0$ is shown as
\begin{multline}
h \left( \mu_{\ell(1)}, \cdots, \mu_{\ell(K)} \right) =
\ifCLASSOPTIONtwocolumn \\ \fi \int_0^{\infty} e^{-\mu_1}
\int_0^{\mu_1} \cdots \int_0^{\mu_{\ell(1)-2}}
\int_0^{\mu_{\ell(1)}} \cdots
\int_0^{\mu_{N-1}} \\
\times \frac{p\left( \mu_{1}, \cdots, \mu_{N} \right)}{g\left(
\mu_{\ell(1)}, \cdots, \mu_{\ell(K)} \right)} \quad d_{\mu_N} \cdots
d_{\mu_{\ell(1)+1}} d_{\mu_{\ell(1)-1}} \cdots d_{\mu_2} d_{\mu_1},
\label{eq:function_h}
\end{multline}
and $h \left( \mu_{\ell(1)}, \cdots, \mu_{\ell(K)} \right)$ for
$\alpha_1>0$ is the same as in (\ref{eq:function_h}) except for the
integrations over $\mu_i$ for $1 \leq i \leq \ell(1)-1$ as well as
$e^{-\mu_1}$ removed. For $\alpha_1 = 0$, $e^{-\mu_1}$ and $\mu_1$
disappear after the integration, while $\mu_1$ is present both in $g
\left( \mu_{\ell(1)}, \cdots, \mu_{\ell(K)} \right)$ and $h \left(
\mu_{\ell(1)}, \cdots, \mu_{\ell(K)} \right)$ for $\alpha_1
> 0$. The introduction in (\ref{eq:Joint_PDF_2}) of $e^{-\mu_1}$ for
$\alpha_1 = 0$ is needed to prevent (\ref{eq:function_h}) from
diverging.

Let's denote
\ifCLASSOPTIONonecolumn
\begin{align}
r \left( \mu_{\ell(1)}, \cdots, \mu_{\ell(K)} \right) = h
\left(\mu_{\ell(1)}, \cdots, \mu_{\ell(K)} \right ) \times g\left(
\mu_{\ell(1)}, \cdots, \mu_{\ell(K)} \right ).
\label{eq:Polynomial_R}
\end{align}
\else
\begin{multline}
r \left( \mu_{\ell(1)}, \cdots, \mu_{\ell(K)} \right) = \\
h \left(\mu_{\ell(1)}, \cdots, \mu_{\ell(K)} \right ) \times g\left(
\mu_{\ell(1)}, \cdots, \mu_{\ell(K)} \right ).
\label{eq:Polynomial_R}
\end{multline}
\fi Then, $r \left( \mu_{\ell(1)}, \cdots, \mu_{\ell(K)} \right)$ is
a polynomial with the smallest degree $(M-Q+1)(N-Q+1)-K$ where $Q$
is an index to indicate the first nonzero $\alpha$, that is, $Q =
\ell(1)$. The proof of this smallest degree is provided in the
Appendix. Since $f \left( \mu_{\ell(1)}, \cdots, \mu_{\ell(K)}
\right ) \leq \hat{f} \left( \mu_{\ell(1)}, \cdots, \mu_{\ell(K)}
\right )$, the right side in (\ref{eq:PEP_nzmin}) is upper bounded
as
\begin{multline}
E \left[ \exp \left(- W \sum\limits_{k=1}^{K} \mu_{\ell{(k)}}
\right) \right] \leq \ifCLASSOPTIONtwocolumn \\ \fi
\int_0^{\infty}
\cdots \int_0^{\mu_{\ell(K-1)}} r \left( \mu_{\ell(1)}, \cdots,
\mu_{\ell(K)} \right ) e^{-(1+W) \sum\limits_{k=1}^K \mu_{\ell(k)}}
\ifCLASSOPTIONtwocolumn
\\ \times
\fi
d_{\mu_{\ell(K)}} \cdots d_{\mu_{\ell(1)}}. \label{eq:PEP_final}
\end{multline}
Note that $1+W \approx W$ for high SNR. In addition, it can be
easily verified that the following equality of a specific term in a
polynomial for $\nu_1 > \nu_2 > \cdots > \nu_K$ holds true;
\ifCLASSOPTIONonecolumn
\begin{align}
\int_0^{\infty} \cdots \int_0^{\nu_{K-1}} \nu_1^{\beta_1} \cdots
\nu_K^{\beta_K} e^{-W \sum\limits_{k=1}^K \nu_{k}} d_{\nu_{K}}
\cdots d_{\nu_{1}} = \zeta W^{ -\left( K + \sum\limits_{k=1}^K
\beta_k \right)} \label{eq:Exponent_calculation}
\end{align}
\else
\begin{multline}
\int_0^{\infty} \cdots \int_0^{\nu_{K-1}} \nu_1^{\beta_1} \cdots
\nu_K^{\beta_K} e^{-W \sum\limits_{k=1}^K \nu_{k}} d_{\nu_{K}}
\cdots d_{\nu_{1}} = \\
\zeta W^{ -\left( K + \sum\limits_{k=1}^K \beta_k \right)}
\label{eq:Exponent_calculation}
\end{multline}
\fi where $\zeta$ is a constant. Since the polynomial $r \left(
\mu_{\ell(1)}, \cdots, \mu_{\ell(K)} \right )$ is a sum of a number
of terms with different degrees, the result of (\ref{eq:PEP_final})
is a sum of the terms of $W$ whose exponent is the corresponding
degree. Furthermore, we are interested in the exponent of $W$ to
figure out the behavior of the diversity, not the exact PEP.
Therefore, we can conclude that PEP is dominated by the term with
the smallest degree of $r \left( \mu_{\ell(1)}, \cdots,
\mu_{\ell(K)} \right )$ which is $(M-Q+1)(N-Q+1)-K$, resulting in
\begin{align}
P\left(\mathbf{c} \rightarrow \mathbf{\hat{c}}\right) &\leq \eta W^{
-(M-Q+1)(N-Q+1)} \nonumber \\
&= \eta \left( \frac{d_{min}^2 \alpha_{nzmin}}{4N} SNR
\right)^{-(M-Q+1)(N-Q+1)} \label{eq:PEP_result}
\end{align}
where $\eta$ is a constant. Fig. \ref{fig:PEP_3x3_S3} shows the
calculations of (\ref{eq:PEP_given_H}) corresponding to several
specific $\alpha$-vectors through Monte-Carlo simulation. Three
dotted straight lines are PEP asymptotes at high SNR whose exponents
correspond to $1$, $4$, and $9$. Regarding the exponent of PEP, we
can see that the calculation of (\ref{eq:PEP_given_H}) through
simulation matches the analysis.

\ifCLASSOPTIONonecolumn
\begin{figure}[!m]
\centering \includegraphics[width =
0.75\linewidth]{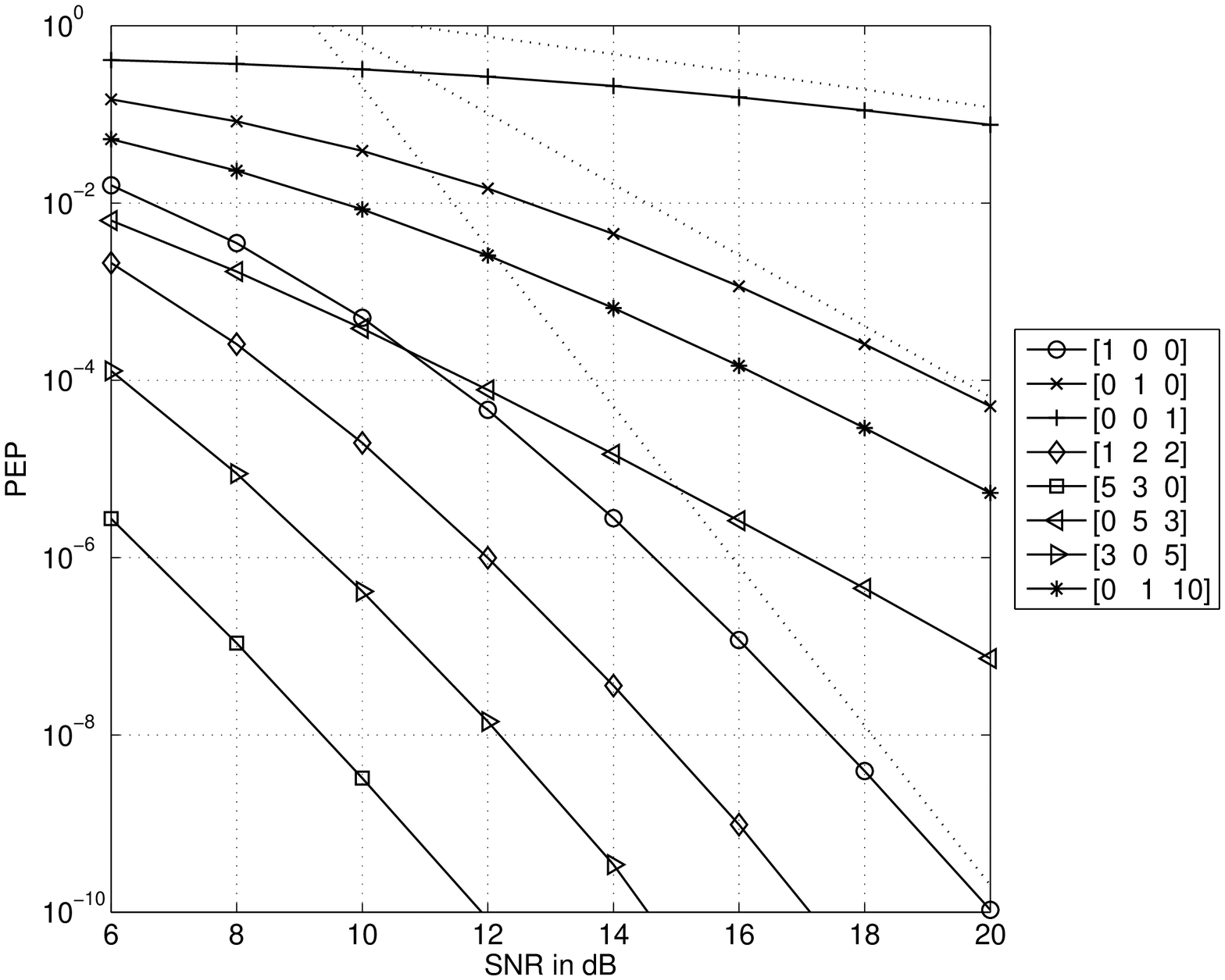} \caption{Monte-Carlo Simulation
Results for $3 \times 3$ $S=3$ case with $4$-QAM}
\label{fig:PEP_3x3_S3}
\end{figure}
\else
\begin{figure}[!t]
\centering \includegraphics[width =
1.1\linewidth]{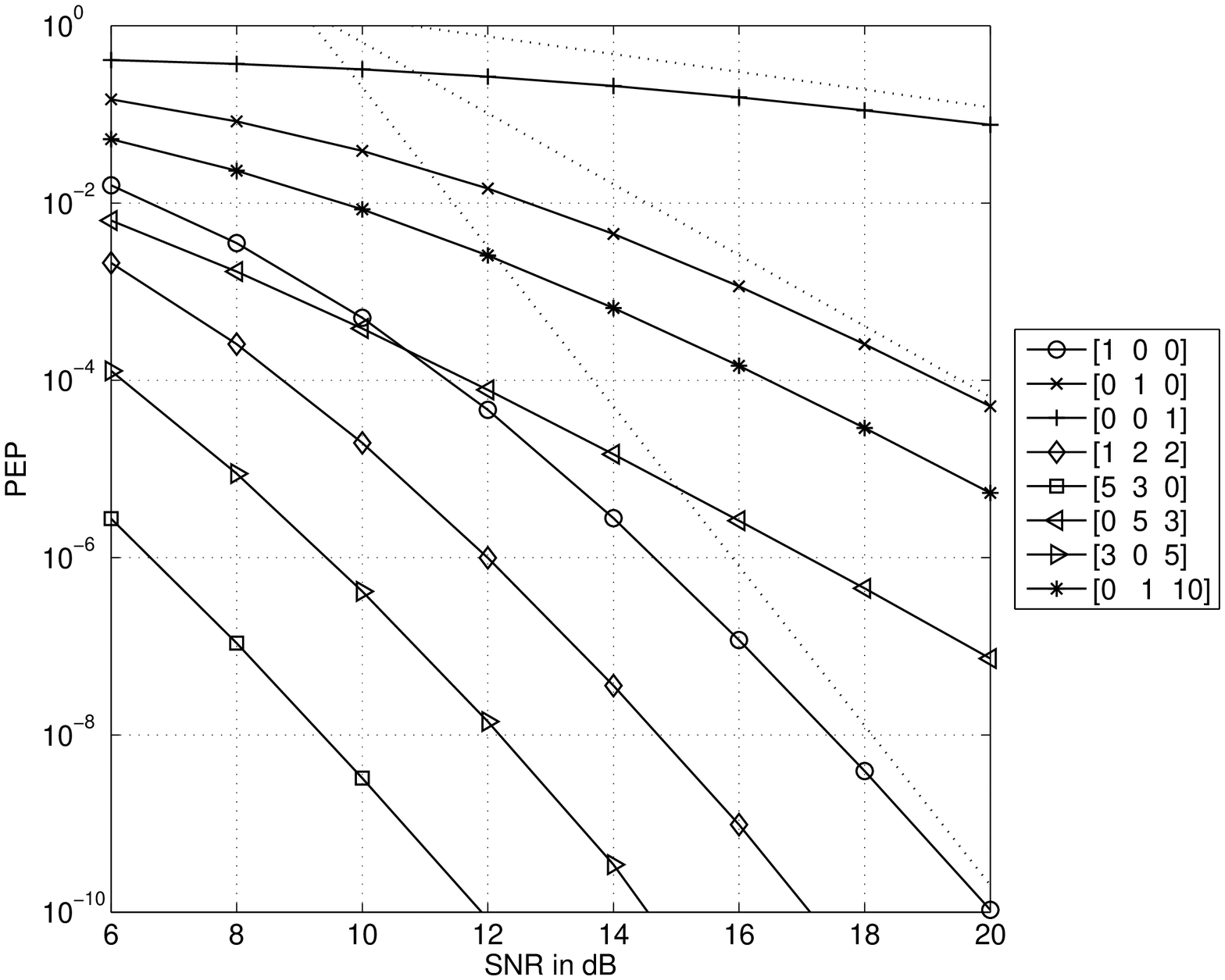} \caption{Monte-Carlo Simulation
Results for $3 \times 3$ $S=3$ case with $4$-QAM}
\label{fig:PEP_3x3_S3}
\end{figure}
\fi

For a rate $k_c / n_c$ binary convolutional code and a fixed
Gray-encoded constellation labeling map in a BICMB system, BER $P_b$
can be bounded as
\begin{align}
P_b \leq \frac{1}{k_c} \sum_{d=d_{free}}^{\infty}
\sum_{i=1}^{W_I(d)} g(d, Q(d, i), \chi) \label{eq:BER_BICMB}
\end{align}
where $g(\cdot)$ is PEP corresponding to each error event, $W_I(d)$
denotes the total input weight of error events at Hamming distance
$d$, and $Q$ is different for each error event. Since BER is
dominated by PEP with the worst exponent term, the diversity order
of a given BICMB system can be represented by
\begin{align}
O_{diversity} = (M-Q_{max}+1)(N-Q_{max}+1) \label{eq:d_order}
\end{align}
where $Q_{max}$ is the maximum $Q$ among the whole set of $Q$'s
corresponding to all of the error events.

\section{Relationship between $Q_{max}$ and code rate} \label{sec:do_bound}

The relationship between $Q_{max}$ and the code with rate $R_c$ is
analyzed by using the same approach as in \cite{KnoppTIT00} which
employes the Singleton bound to calculate the minimum distance of a
non-binary block code. Let's define $d_{E,s}(\mathbf{c},
\mathbf{\hat{c}})$ as the Euclidean distance between the mapped
symbols of the two codewords residing on the $s^{th}$ subchannel,
$d_{E,s}(\mathbf{c}, \mathbf{\hat{c}}) = \sum_{k=1}^L | y_{k,s} -
\hat{y}_{k,s} |^2$ where $y_{k,s}$ and $\hat{y}_{k,s}$ are symbols
on the $s^{th}$ subchannel at the time index $k$ from the codeword
$\mathbf{c}$ and $\mathbf{\hat{c}}$, respectively. If $\alpha_s$ is
equal to zero, then all of the coded bits on the $s^{th}$ subchannel
of the two codewords are the same. Since we assume that the
consecutive bits are mapped over different symbols, the symbols
corresponding to the same coded bits of the $s^{th}$ subchannel are
also the same, resulting in $d_{E,s}(\mathbf{c}, \mathbf{\hat{c}}) =
0$. Then, the parameter $Q$ can be viewed as an index to the first
non-zero element in a vector $[d_{E,1}(\mathbf{c}, \mathbf{\hat{c}})
\quad d_{E,2}(\mathbf{c}, \mathbf{\hat{c}}) \quad \cdots \quad
d_{E,S}(\mathbf{c}, \mathbf{\hat{c}})]$. In the case of a pair of
the codewords that has $S-1$ non-zero $d_{E,s}(\mathbf{c},
\mathbf{\hat{c}})$'s, $Q$ can be $2$ because of the vector type $[0
\, \times \, \times \, \cdots \, \times]$, or $1$ from $[\times \, 0
\, \times \, \times \, \cdots \, \times]$, $[\times \, \times \, 0
\, \times \, \times \, \cdots \, \times]$, $\cdots$, $[\times \,
\times \, \times \, \cdots \, \times \, 0]$, where $\times$ stands
for non-zero value. In general, for a pair of the codewords that has
$\delta_H$ non-zero $d_{E,s}(\mathbf{c}, \mathbf{\hat{c}})$'s, $Q$
is bounded as
\begin{align}
Q \leq S - \delta_H + 1.
\label{eq:Q_and nonzero_EucDist}
\end{align}

If we consider the $L$ symbols transmitted on each subchannel as a
super-symbol over $\chi^L$, then the transmitted symbols for all the
subchannels in a block can be viewed as a vector of length $S$
super-symbols. For convenience, we will call this vector of
super-symbols as a symbol-wise codeword. We will now introduce a
distance between $\mathbf{c}$ and $\mathbf{\hat{c}}$, which we will
call $\delta_H$, as the number of non-zero elements in the vector
$[d_{E,1}(\mathbf{c}, \mathbf{\hat{c}}) \quad d_{E,2}(\mathbf{c},
\mathbf{\hat{c}}) \quad \cdots \quad d_{E,S}(\mathbf{c},
\mathbf{\hat{c}})]$. This distance is similar to the Hamming
distance but it is between two non-binary symbol-wise codewords. By
using the Singleton bound which is also applicable to non-binary
codes, we can calculate the minimum distance of the symbol-wise
codewords in a way similar to finding the minimum Hamming distance
of binary codes. Let's define $\mathcal{M}$ as the number of
distinct symbol-wise codewords. Then we can see that $\mathcal{M} =
2^{mLSR_c}$ from Fig. \ref{fig:eq_bicmb}. Let $k$ $(0 < k \leq S-1)$
denote the integer value satisfying $2^{mL(k-1)} < \mathcal{M} \leq
2^{mLk}$. Since $\mathcal{M} > 2^{mL(k-1)}$, there necessarily exist
two symbol-wise codewords whose $k-1$ elements are the same. From
the Singleton bound \cite{SingletonTIT64}, the minimum distance of
these symbol-wise codewords $\delta_{H,min}$ is expressed as
$\delta_{H,min} \leq S - k + 1$. Since $2^{mLSR_c} \leq
2^{mL(S-\delta_{H,min}+1)}$, we get
\begin{align}
\delta_{H,min} \leq S - \lceil S \cdot R_c \rceil + 1
\label{eq:Singleton_bound}
\end{align}
using the fact that $\delta_{H,min}$ is an integer value.

For a given BICMB system with $\delta_{H,min}$, it is true that the
distance $\delta_H$ between any pair of the codewords  is always
larger than or equal to the minimum distance $\delta_{H,min}$. By
combining the inequalities of $\delta_{H} \geq \delta_{H,min}$ and
(\ref{eq:Q_and nonzero_EucDist}), we get $\delta_{H,min} \leq
\delta_{H} \leq S - Q + 1$, leading to the following inequality as
\begin{align}
Q \leq S - \delta_{H,min} + 1. \label{eq:Q_and_H_d_min}
\end{align}
From the inequality (\ref{eq:Q_and_H_d_min}), the maximum $Q$ among
the whole set of $Q$'s can be found as
\begin{align}
Q_{max} = S - \delta_{H,min} + 1. \label{eq:Q_max_and_H_d_min}
\end{align}
The inequality (\ref{eq:Singleton_bound}) and the equation
(\ref{eq:Q_max_and_H_d_min}) result in the following inequality as
\begin{align}
Q_{max} \geq \lceil S \cdot R_c \rceil. \label{eq:Q_max_and_rate}
\end{align}
The relationship (\ref{eq:Q_max_and_rate}) can be supported by the
examples in Section \ref{sec:alpha_Spectra} where the $1/2$-rate
convolutional code is used in $S=3$ BICMB system with different
spatial de-multiplexers. Since $Q_{max}$ can be $2$ or $3$ according
to (\ref{eq:Q_max_and_rate}), the rotating spatial de-multiplexer
used to calculate $\mathbf{T}_2$ in (\ref{eq:Transfunc_S3R1_2})
makes $Q_{max}$ equal to $2$ while that of $\mathbf{T}_3$ in
(\ref{eq:Transfunc_S3R1_2_diffdemux}) makes $Q_{max}$ equal to $3$.
By considering the calculated diversity order of (\ref{eq:d_order}),
the maximum diversity order for a given code rate BICMB system is
achieved by choosing the convolutional encoder and the spatial
de-multiplexer satisfying $Q_{max} = \lceil S \cdot R_c \rceil$. In
this case, the maximum achievable diversity order is
\begin{align}
O_{diversity} = (M-\lceil S \cdot R_c \rceil+1)(N-\lceil S \cdot R_c
\rceil+1). \label{eq:d_order_R_c}
\end{align}
Based on (\ref{eq:d_order_R_c}), Fig. \ref{fig:S_vs_Rc} depicts the
relationship between the code rate $R_c$, the number of streams $S$,
and the maximum achievable diversity order. The whole combinations
of $S$ and $R_c$ are divided into the four regions each representing
the maximum achievable diversity order. For example, such
combinations as $(S$, $R_c)$ $=$ $(2$, $1/2)$, $(2$, $1/3)$, $(2$,
$1/4)$, $(3$, $1/3)$, $(3$, $1/4)$ in the region with the legend of
$MN$ achieve the full diversity order of $MN$.
\ifCLASSOPTIONonecolumn
\begin{figure}[!m]
\centering \includegraphics[width = 0.6\linewidth]{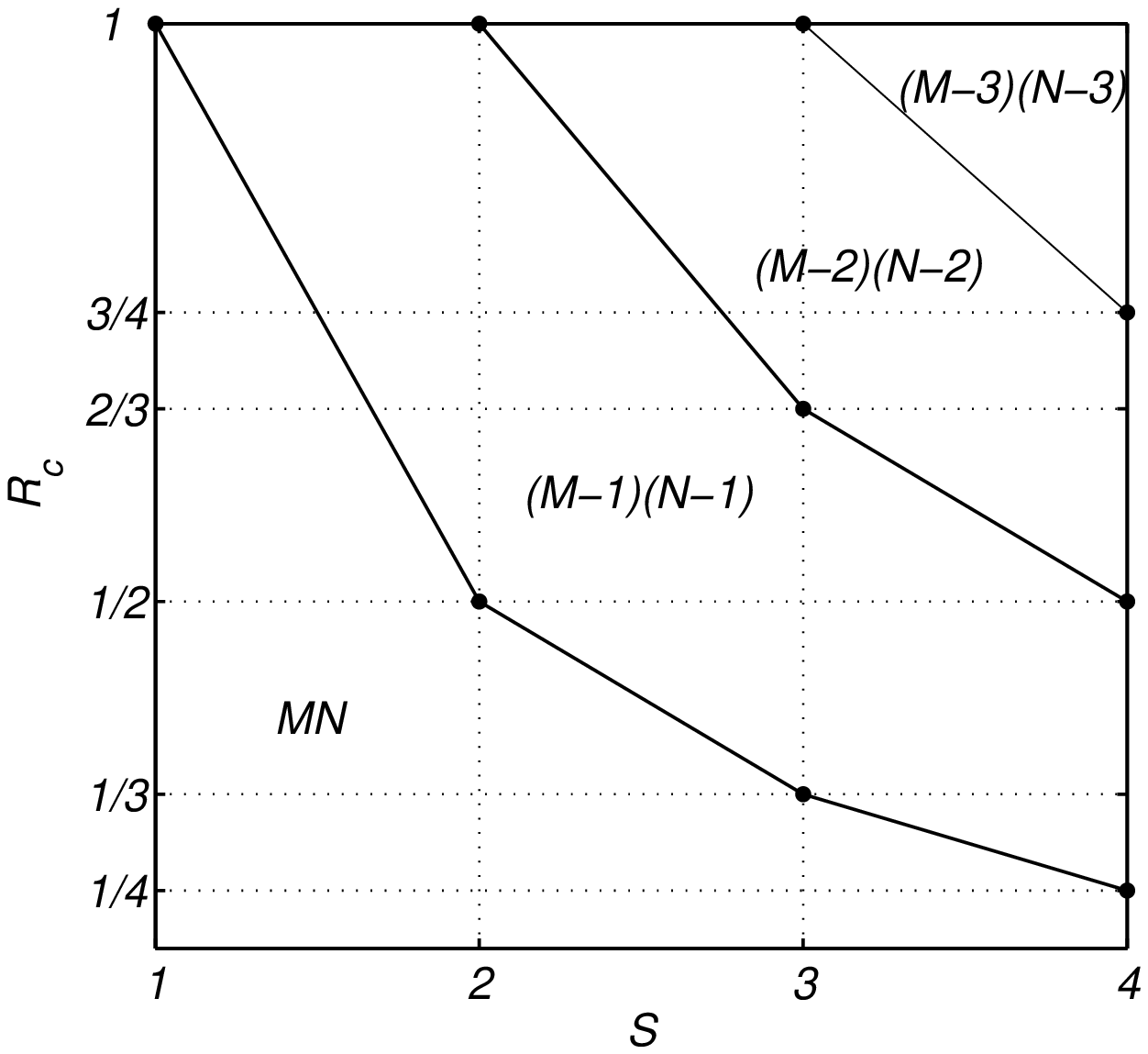}
\caption{Multiplexing-Diversity-Rate Relationship}
\label{fig:S_vs_Rc}
\end{figure}
\else
\begin{figure}[!t]
\centering \includegraphics[width = 1.0\linewidth]{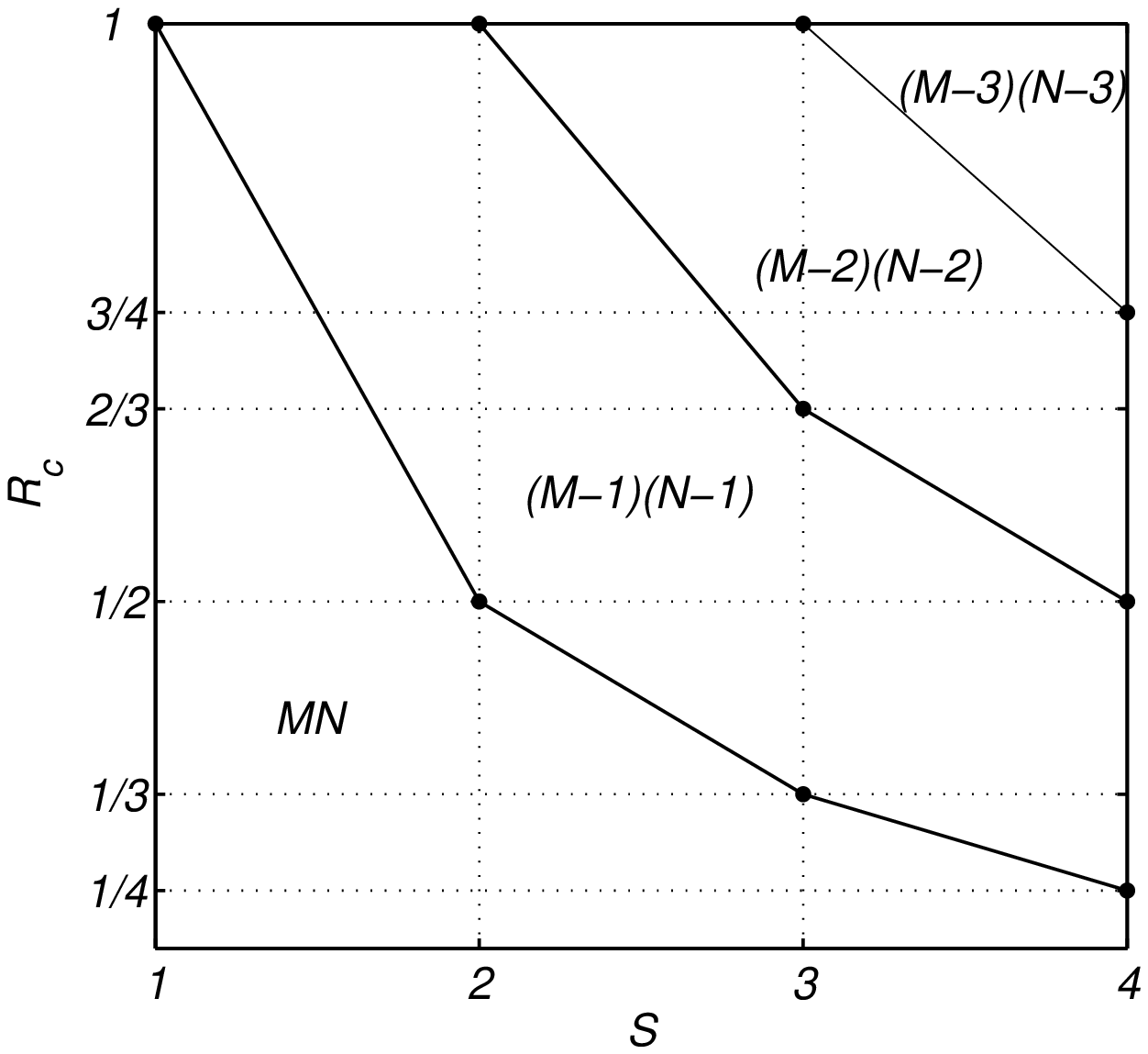}
\caption{Multiplexing-Diversity-Rate Relationship}
\label{fig:S_vs_Rc}
\end{figure}
\fi

Since we assumed in the previous description that there exist the
convolutional encoder and the spatial de-multiplexer which satisfy
the relation \mbox{$Q_{max} = \lceil S \cdot R_c \rceil$}, we will
show the specific design method of the interleaver from a given
convolutional encoder to ensure the relation. The following method
is not the unique solution to guarantee the maximum achievable
condition, but simple to state the concept. Let's consider a BICMB
system with $S$ subchannels and the code rate $R_c = k_c/n_c$
convolutional code. Each of \mbox{$P = LCM(n_c, S)$} coded bits is
distributed to the $S$ streams in the order specified by the
interleaving pattern. Since each stream needs to be evenly employed
for a period, $P/S$ coded bits are assigned on each stream. To
guarantee \mbox{$Q_{max} = \lceil S \cdot R_c \rceil$}, it is
sufficient to consider only the first branches that split from the
zero state in one period because of the repetition property of the
convolutional code. We incorporate the basic idea that once the
$s^{th}$ stream is assigned to an error bit of the first branch,
obviously, all of the error events containing that branch give
$\alpha_s > 0$, resulting in $Q \leq s$. By extending this idea, we
can summarize the assignment procedure as
\begin{enumerate}
\item the lowest available subchannel is assigned to the error bit
position of one of the first branches which have not yet assigned to
any subchannel,
\item the procedure 1) is repeated until all of the first branches
are assigned to one of the subchannels. If all of the first branches
are assigned to one of the subchannels, the assignment procedure
quits after the rest of subchannels are assigned randomly to the
unassigned bit positions, subject to satisfying the rate condition
on each subchannel.
\end{enumerate}

We will explain the procedure above by using the example of Fig.
\ref{fig:trellis_alpha} where $P = 4$ and the number of available
assignment for each subchannel is $P/S = 1$. From the trellis, we
can see there are $2$ first branches that split from the zero state
for one period. According to the procedure above, we need to assign
the best subchannel to one of the first branches. In this example,
let's assign it to the dummy variable $a$. This ensures $\alpha_1 >
0$ for all the error events stemming from this branch, resulting in
$Q = 1$. For the second branch connecting $X_i$ to $X_{11}$, we need
to assign the next available lowest subchannel, which is $2$, to the
dummy variable $c$. As a result, we can see that $Q \leq 2$ for the
error events that share this branch. Since all of the first two
branches are assigned, the unassigned dummy variables are allocated
randomly with $3$ and $4$. This procedure assures that $Q_{max}$ is
equal to $1$ or $2$ for this BICMB system. On the other hand,
$Q_{max} \geq 2$ from the equation (\ref{eq:Q_max_and_rate})
resulting from the Singleton bound. Therefore, this method
guarantees $Q_{max} = 2$, which is the condition to achieve the
maximum diversity order for the given convolutional code.

\section{Simulation Results} \label{sec:result}

To show the validity of the diversity order analysis in Section
\ref{sec:cal_order} using the parameter $Q_{max}$ calculated by the
method in Section \ref{sec:alpha_Spectra}, BER against SNR are
derived through a Monte-Carlo simulation. Fig.
\ref{fig:3x3_S3_diffstr} shows BER performances for the cases
corresponding to $\mathbf{T}_1$, $\mathbf{T}_2$, $\mathbf{T}_3$ in
(\ref{eq:Transfunc_S2R1_2}), (\ref{eq:Transfunc_S3R1_2}), and
(\ref{eq:Transfunc_S3R1_2_diffdemux}). The well-known reference
curves achieving the full diversity order of $MN$ are drawn from the
Alamouti code for the $2 \times 2$ case and $1/2$-rate orthogonal
space-time block code (OSTBC) for the $3 \times 3$ case
\cite{TarokhJSAIC99}. From (\ref{eq:Transfunc_S2R1_2}), $Q_{max}$
for $\mathbf{T}_1$ is found to be $1$ because $\alpha_s \geq 1$ for
$s=1,2$ in all of the $\alpha$-vectors. In this case, as predicted
by the analysis in \cite{akayTC06BICMB}, \cite{akayTC06BICMB_arxiv},
the diversity order equals $4$ by calculating (\ref{eq:d_order})
with $M=N=2$. From the figure, we can see that BER curve for
$\mathbf{T}_1$ is parallel to that of $2 \times 2$ Alamouti code.
Since $Q_{max}$ for $\mathbf{T}_2$ is $2$ due to the vector $[0\,
5\, 3]$, the calculated diversity order is $4$ in the case of
$M=N=S=3$. This can be verified by Fig. \ref{fig:3x3_S3_diffstr},
losing the full diversity order $9$. Although the same number of
subchannels and the same convolutional code as for $\mathbf{T_2}$
are used, the different spatial de-multiplexer from that of
$\mathbf{T_2}$, described in Section \ref{sec:alpha_Spectra} for
$\mathbf{T_3}$, gives no diversity gain at all. The reason for this
is that the vector $[0\, 0\, 5]$ which can be observed from the
transfer function in $\mathbf{T}_3$ makes $Q_{max}$ $3$ resulting in
the calculated diversity order of $1$ in the equation
(\ref{eq:d_order}) with $M=N=S=3$. This matches the simulation
result.

\ifCLASSOPTIONonecolumn
\begin{figure}[!m]
\centering \includegraphics[width
=0.6\linewidth]{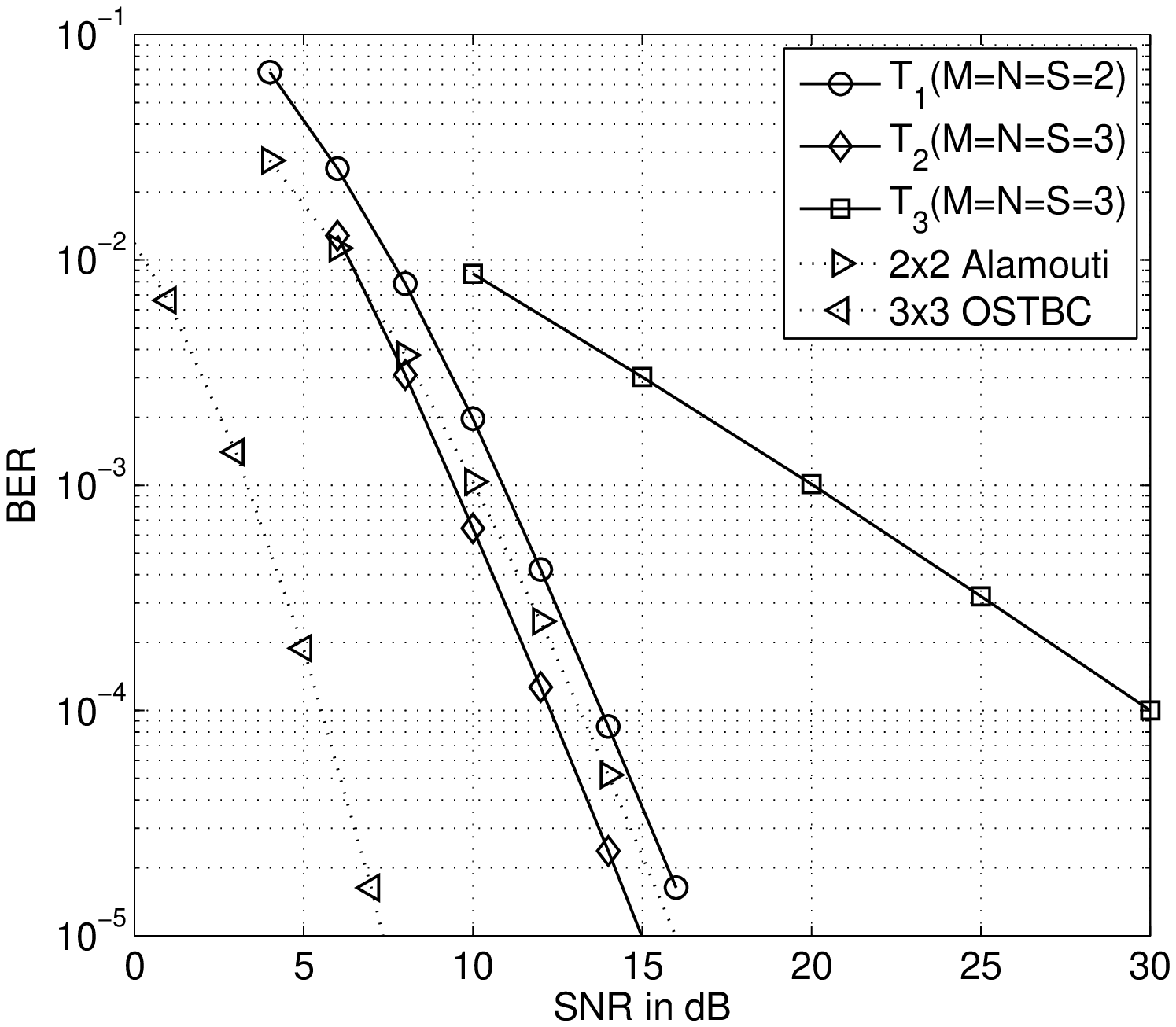} \caption{Simulation results for
a $4$-state $1/2$-rate convolutional code with different spatial
de-multiplexers. $4$-QAM is used for all of the curves.}
\label{fig:3x3_S3_diffstr}
\end{figure}
\else
\begin{figure}[!t]
\centering \includegraphics[width = 1\linewidth]{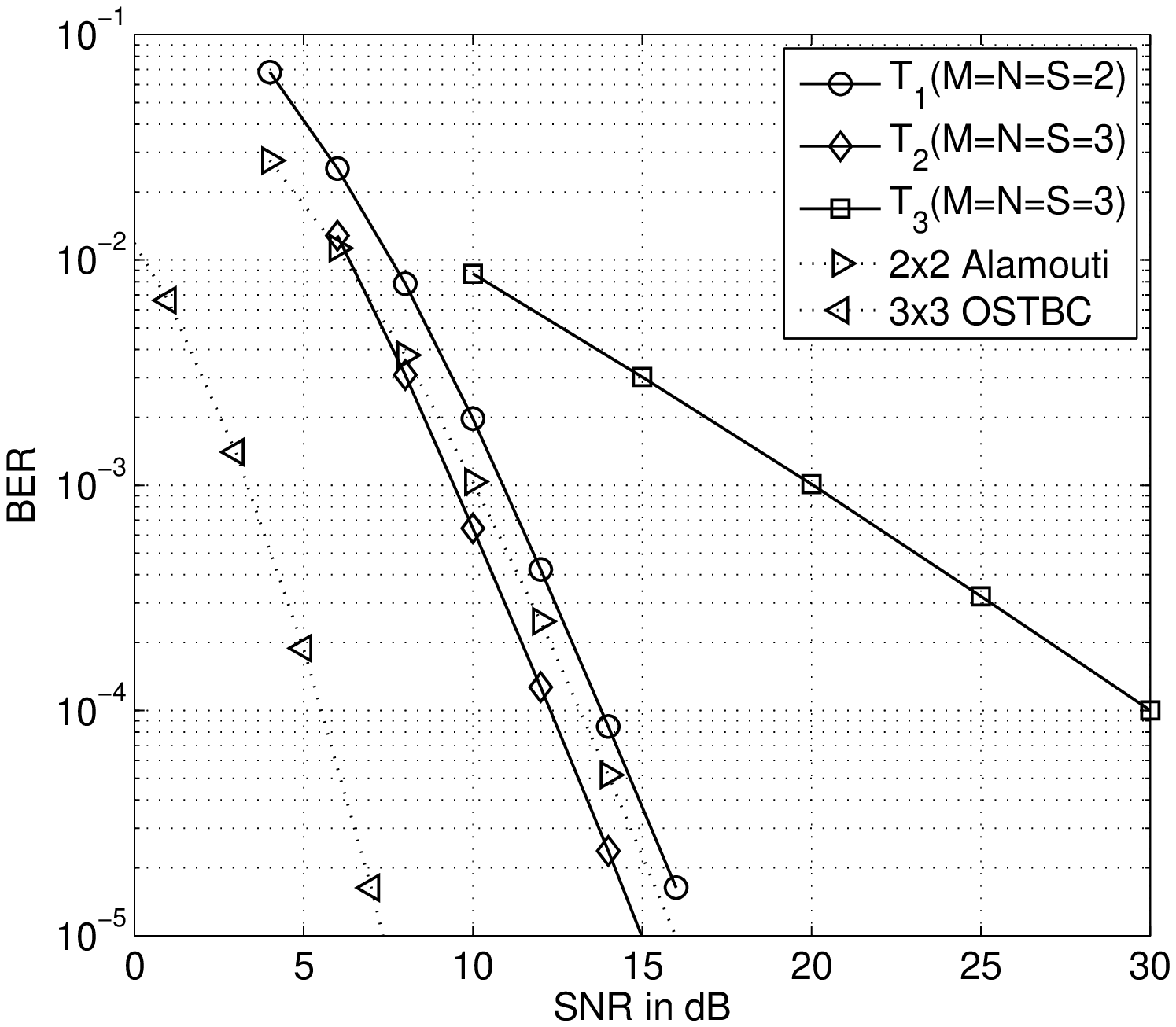}
\caption{Simulation results for a $4$-state $1/2$-rate convolutional
code with different spatial de-multiplexers. $4$-QAM is used for all
of the curves.} \label{fig:3x3_S3_diffstr}
\end{figure}
\fi

Table \ref{tab:alpha_vectors} shows results of a computer search of
the $\alpha$-vectors of BICMB with industry standard $64$-state
convolutional codes and a simple rotating spatial de-multiplexer.
The generator polynomials for rates $1/2$ and $1/3$ are $(133, 171)$
and $(133, 145, 175)$ in octal, respectively. For the high rate
codes such as $2/3$ and $3/4$, the perforation matrices in
\cite{beginTCOM90} are used from the $1/2$-rate original code.
Instead of displaying the whole transfer functions, we present only
three $\alpha$-vectors among such a number of dominant
$\alpha$-vectors that lead to $Q_{max}$. The search results comply
with the bound \mbox{$Q_{max} \geq \lceil S \cdot R_c \rceil$} in
(\ref{eq:Q_max_and_rate}) as was analyzed in Section
\ref{sec:do_bound}.

\ifCLASSOPTIONonecolumn
\begin{table}[!m]
\else
\begin{table}[!t]
\fi \caption{Search Results of the dominant $\alpha$-vectors for
$64$-state convolutional codes}
\begin{center}
\begin{tabular}{|c|c|c|c|c|}
\hline
$S$ & rate & $d_{free}$ & dominant $\alpha$-vectors & $Q_{max}$ \\
\hline \hline
\multirow{3}*{$2$} & $1/2$ & $10$ & $[3\,\, 7]$ $[4\,\, 6]$ $[5\,\, 5]$ & $1$ \\
\cline{2-5}
& $2/3$ & $6$ & $[0\,\, 12]$ $[0\,\, 14]$ $[0\,\, 15]$ & $2$ \\
\cline{2-5}
& $3/4$ & $5$ & $[0\,\, 8]$ $[0\,\, 10]$ $[0\,\, 12]$ & $2$ \\
\hline \hline
\multirow{4}*{$3$} & $1/3$ & $15$ & $[3\,\, 6\,\, 6]$ $[5\,\, 4\,\, 6]$ $[4\,\, 6\,\, 6]$ & $1$ \\
\cline{2-5}
& $1/2$ & $10$ & $[0\,\, 7\,\, 7]$ $[0\,\, 8\,\, 6]$ $[0\,\, 9\,\, 7]$ & $2$ \\
\cline{2-5}
& $2/3$ & $6$ & $[0\,\, 4\,\, 5]$ $[0\,\, 6\,\, 3]$ $[0\,\, 4\,\, 6]$ & $2$ \\
\cline{2-5}
& $3/4$ & $5$ & $[0\,\, 0\,\, 13]$ $[0\,\, 0\,\, 15]$ $[0\,\, 0\,\, 17]$& $3$ \\
\hline
\end{tabular}
\end{center}
\label{tab:alpha_vectors}
\end{table}

Fig. \ref{fig:2x2} shows the BER performance of the $2 \times 2$
$S=2$ BICMB system with the $64$-state convolutional code and a
simple rotating spatial de-multiplexer. The diversity orders of the
systems with punctured codes are $1$ because both $Q_{max}$ values
corresponding to the codes shown in Table \ref{tab:alpha_vectors}
are $2$, while the system with the $1/2$-rate convolutional code,
whose $Q_{max}$ is equal to $1$, achieves full diversity order of
$4$.

\ifCLASSOPTIONonecolumn
\begin{figure}[!m]
\centering
\includegraphics[width = 0.6\linewidth]{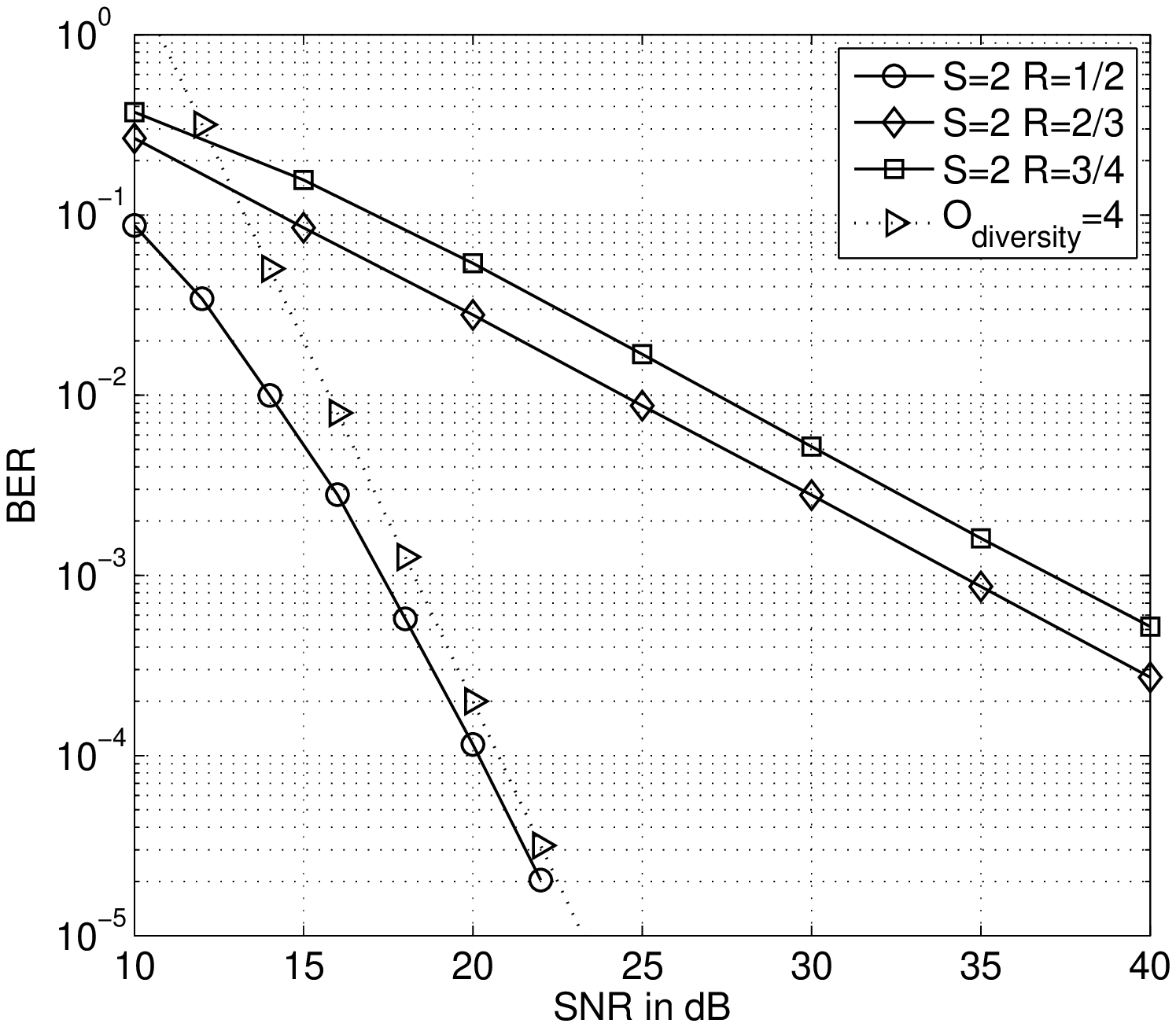}
\caption{Simulation results for the $2 \times 2$ case where $16$-QAM
is used for all of the curves} \label{fig:2x2}
\end{figure}
\else
\begin{figure}[!t]
\centering
\includegraphics[width = 1\linewidth]{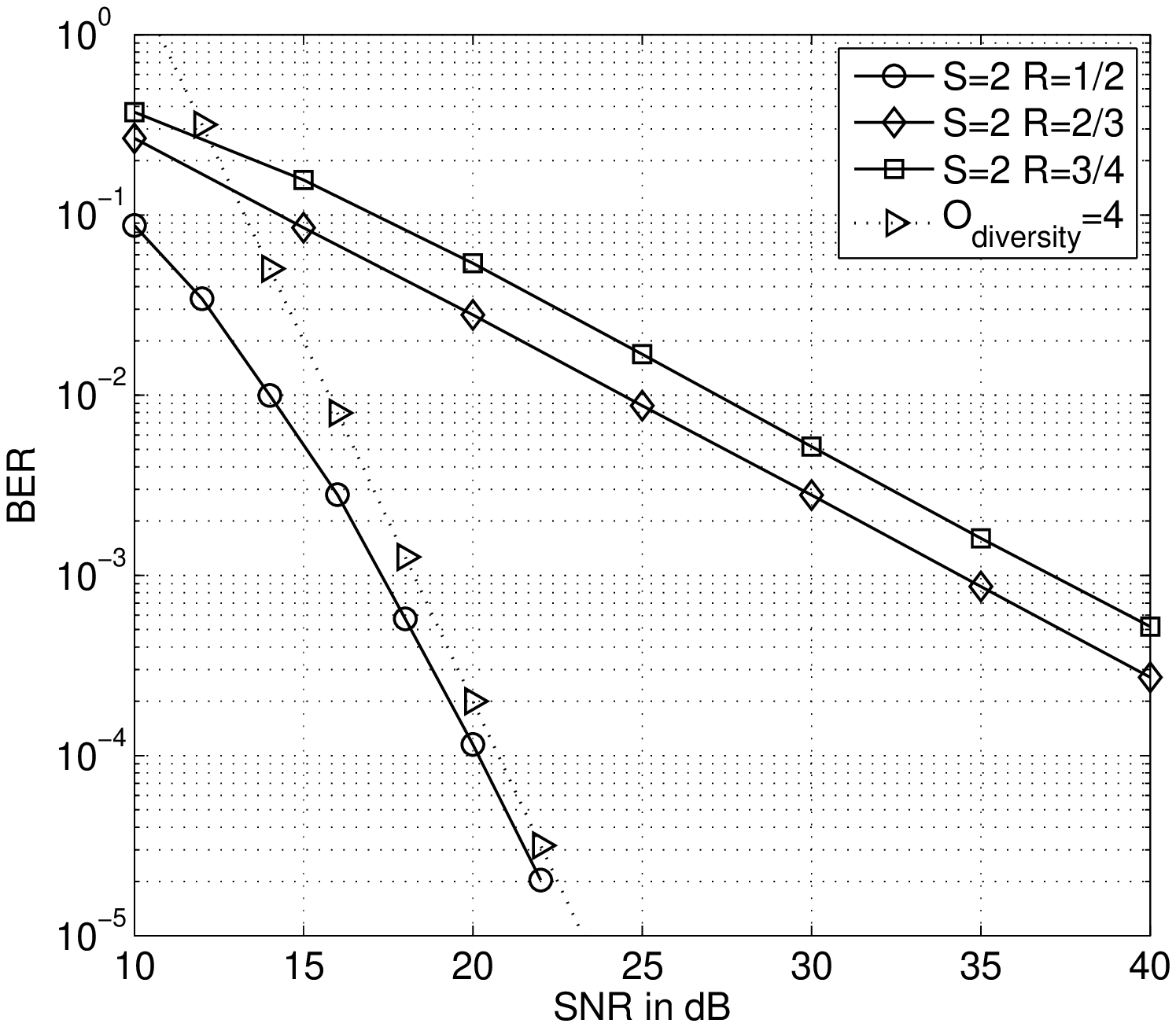}
\caption{Simulation results for the $2 \times 2$ case where $16$-QAM
is used for all of the curves} \label{fig:2x2}
\end{figure}
\fi

As shown in Fig. \ref{fig:3x3_S3}, for a $3 \times 3$ system with
$3$ streams, only $1/3$-rate convolutional code achieves full
diversity order of $9$ since other codes have $Q_{max}$ of larger
than $1$ as given in Table \ref{tab:alpha_vectors}. The analytically
calculated diversity orders by using (\ref{eq:d_order}) and Table
\ref{tab:alpha_vectors} are $4$, $4$, $1$ for $1/2$, $2/3$, $3/4$
respectively, which can be easily verified from Fig.
\ref{fig:3x3_S3} by being compared with the asymptotes. For the
rate-$3/4$ code with the same spatial de-multiplexer, reducing one
stream improves the performance dramatically. The diversity order of
this case is $4$ from the equation (\ref{eq:d_order}) with $M=N=3$
and $Q_{max}=2$.

\ifCLASSOPTIONonecolumn
\begin{figure}[!m]
\centering \includegraphics[width = 0.6\linewidth]{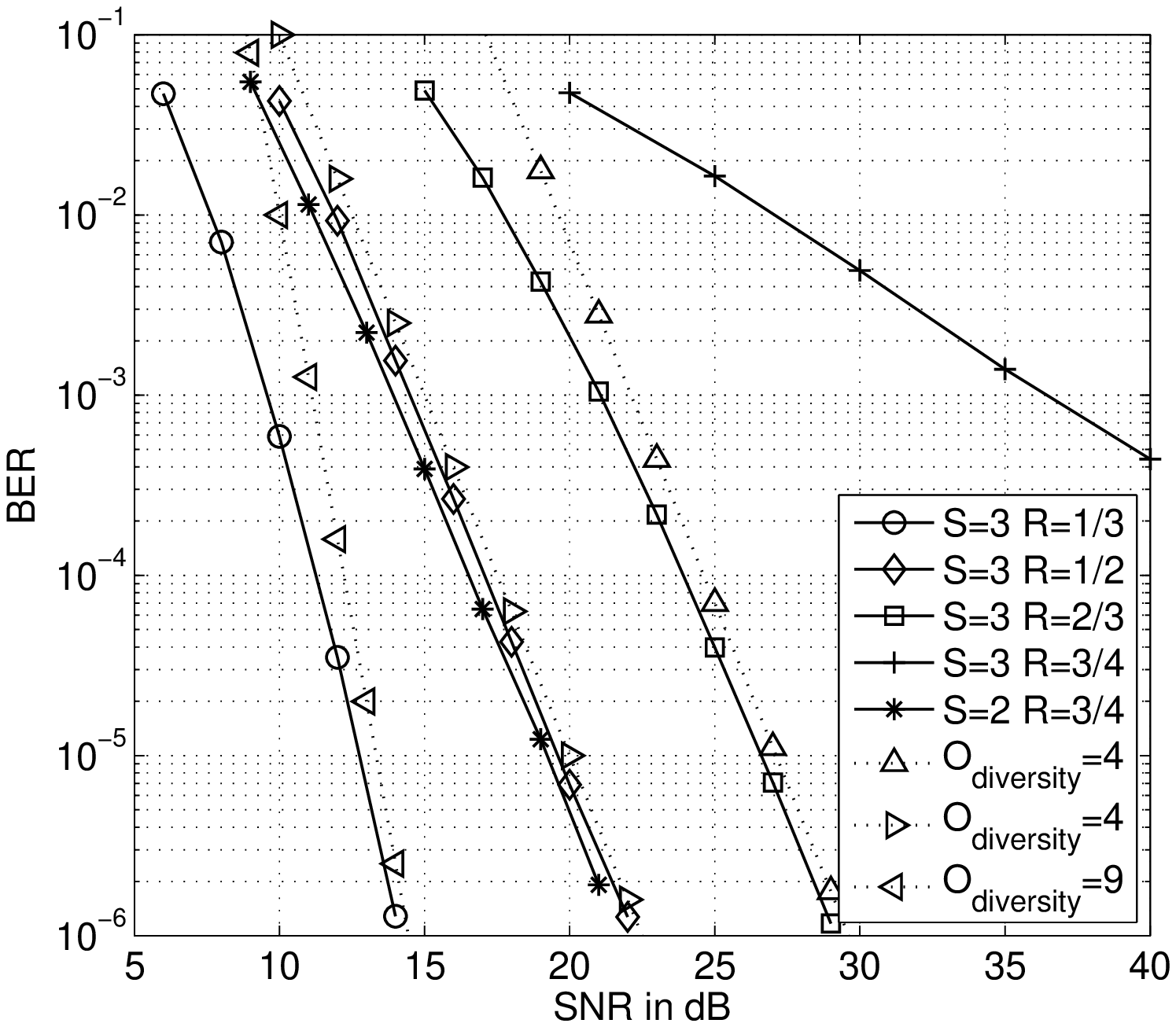}
\caption{Simulation results for $3 \times 3$ case where $16$-QAM is
used for all of the curves} \label{fig:3x3_S3}
\end{figure}
\else
\begin{figure}[!t]
\centering \includegraphics[width = 1\linewidth]{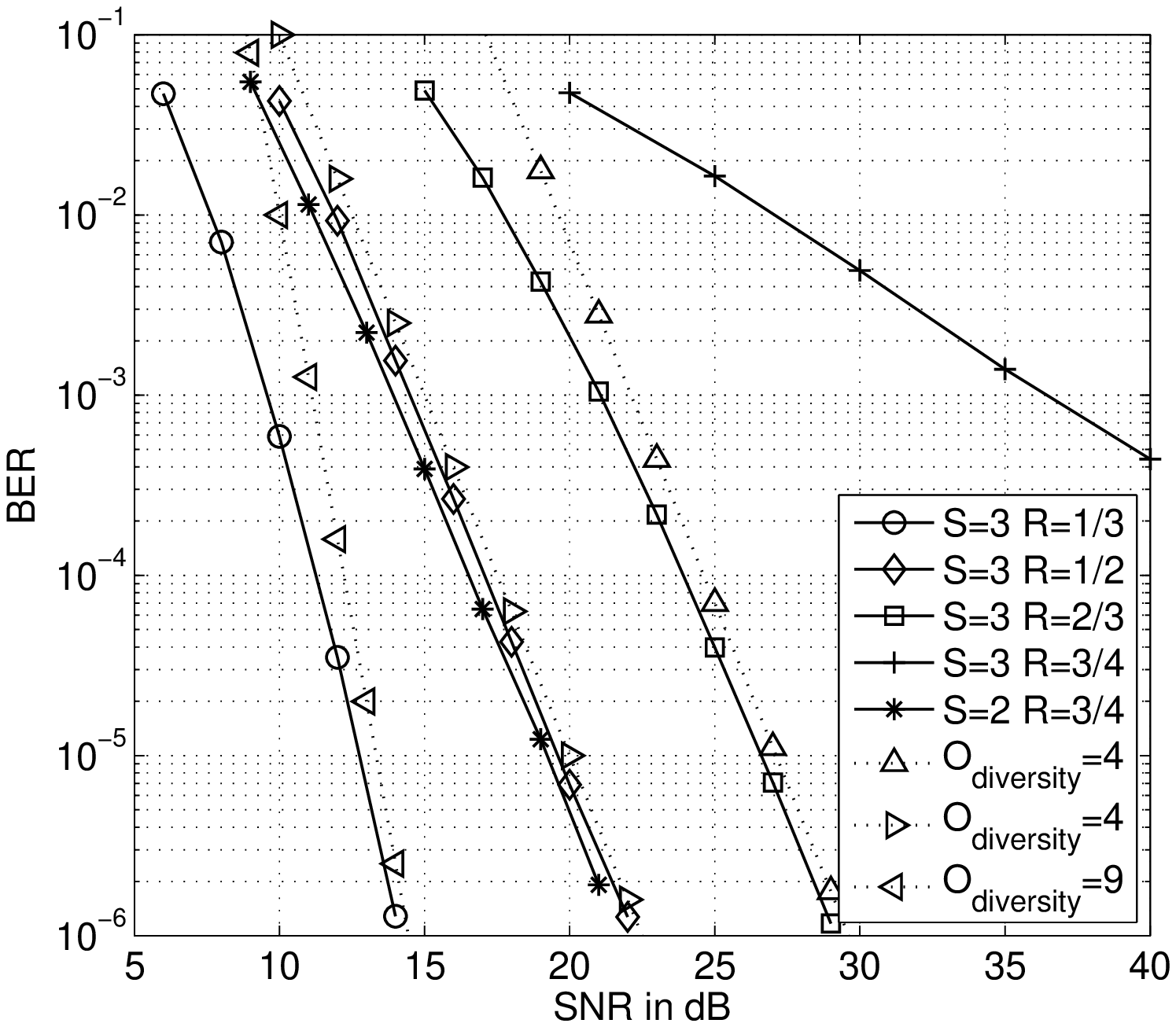}
\caption{Simulation results for $3 \times 3$ case where $16$-QAM is
used for all of the curves} \label{fig:3x3_S3}
\end{figure}
\fi

\section{Conclusion} \label{sec:conclusion}

In this paper, we investigated the diversity order of BICMB when the
interleaver does not meet the previously introduced design criteria.
We introduced a method to calculate the $\alpha$-vectors from a
given convolutional code and a spatial de-multiplexer by using a
transfer function. By using this method, the $\alpha$-vectors that
do not fulfill the full diversity order criteria are quantified.
Then, the diversity behavior corresponding to the $\alpha$-vectors
was analyzed through PEP calculation. The exponent of PEP between
two codewords is $(M-Q+1)(N-Q+1)$ where $Q$ is the first index to
the non-zero element in the $\alpha$-vector. Since BER is dominated
by PEP with the smallest exponent, the diversity order is
$(M-Q_{max}+1)(N-Q_{max}+1)$, where $Q_{max}$ is the maximum among
$Q$'s corresponding to each $\alpha$-vector. We provided the
simulation results that verify the analysis. We also showed that
$Q_{max}$ is lower bounded by the product of the code rate and the
number of streams. This result indicated that we need $R_c \leq 1/S$
to achieve the full diversity order of $NM$.

\appendix[Proof of the smallest degree]

Since $r \left( \mu_{\ell(1)}, \cdots, \mu_{\ell(K)} \right)$ is a
product of the two polynomials as shown in (\ref{eq:Polynomial_R}),
the smallest degree of $r \left( \mu_{\ell(1)}, \cdots,
\mu_{\ell(K)} \right)$ is a sum of the smallest degrees of each
polynomial. Let's denote $D_{g, smallest}$ as the smallest degree of
the polynomial $g \left( \mu_{\ell(1)}, \cdots, \mu_{\ell(K)}
\right)$. It is easily found that all of the terms in
(\ref{eq:function_g}) have the same degree. Therefore,
\begin{align}
D_{g,smallest} = K(M-N)+K(K-1) \label{eq:D_g_smallest}
\end{align}
where the degree of $K(M-N)$ is contributed by the $K$ factors of
the form $\mu_{\ell(k)}^{M-N}$, and $K(K-1)$ comes from the
$\binom{K}{2}$ factors in the form of $\left( \mu_{\ell(k)} -
\mu_{\ell(j)} \right)^2$.

To calculate the smallest degree of the polynomial $h \left(
\mu_{\ell(1)}, \cdots, \mu_{\ell(K)} \right)$, we first focus on the
case of $\alpha_1 = 0$. The polynomial $p\left( \mu_{1}, \cdots,
\mu_{N} \right)$ in (\ref{eq:Polynomial_p}) has $N$ factors of the
form $\mu_i^{M-N}$ and $\binom{N}{2}$ factors of the form $\left(
\mu_i - \mu_j \right)^2$. The division by $g\left( \mu_{\ell(1)},
\cdots, \mu_{\ell(K)} \right)$ makes the common factors eliminated,
leaving $N-K$ and $\binom{N}{2} - \binom{K}{2}$ factors of the form
$\mu_i^{M-N}$ and $\left( \mu_i - \mu_j \right)^2$, respectively.
Hence, the resulting polynomial $p\left( \mu_{1}, \cdots, \mu_{N}
\right) / g\left( \mu_{\ell(1)}, \cdots, \mu_{\ell(K)} \right)$ has
degree
\begin{align}
D_{h,org} = (N-K)(M-N)+N(N-1)-K(K-1). \label{eq:D_org}
\end{align}

The integration over $\mu_i$ for $1 \leq i \leq \ell(1)-1$ in
(\ref{eq:function_h}) makes the variables $\mu_i$, $1 \leq i \leq
\ell(1)-1$ vanish. Since all the terms in $p\left( \mu_{1}, \cdots,
\mu_{N} \right) / g\left( \mu_{\ell(1)}, \cdots, \mu_{\ell(K)}
\right)$ have different distributions on the degrees of the
individual variables although they have the same degree as an entire
term, the smallest degree of $h\left( \mu_{\ell(1)}, \cdots,
\mu_{\ell(K)} \right)$ is determined by the term which has the
largest degree of the vanishing variables of $p\left( \mu_{1},
\cdots, \mu_{N} \right) / g\left( \mu_{\ell(1)}, \cdots,
\mu_{\ell(K)} \right)$. It's not necessary to find all the terms
with the largest degree of the vanishing variables. Instead, we can
see that one of those terms, whose degree is $D_{h,org}$, includes
the following factors
\begin{align}
\prod\limits_{i=1}^{\ell(1)-1} \mu_{i}^{M-N} \prod\limits_{j>i}^{N}
\mu_i^2. \label{eq:largest_term}
\end{align}
In this case, the degree for the vanishing variables in
(\ref{eq:largest_term}) is
\ifCLASSOPTIONonecolumn
\begin{align}
D_{h,vanishing} = (\ell(1)-1)(M-N)+2N(\ell(1)-1)-\ell(1)(\ell(1)-1)
\label{eq:D_vanished}
\end{align}
\else
\begin{multline}
D_{h,vanishing} = (\ell(1)-1)(M-N) \\
+2N(\ell(1)-1)-\ell(1)(\ell(1)-1) \label{eq:D_vanished}
\end{multline}
\fi
where $(\ell(1)-1)(M-N)$ is contributed by the $\ell(1)-1$
factors of the form $\mu_i^{M-N}$, and the rest of the degrees are
calculated from the factors of the form $\mu_i^2$.

Finally, the integration over $\mu_i$ for $\ell(1)+1 \leq i \leq N$
accumulates the degree of the current variables and adds up to the
degree of the corresponding $\mu_k, k \in \Upsilon$, where an
ordered set $\Upsilon$ is defined as $\left\{i:\alpha_i>0 \,
\textrm{for }1 \leq i \leq S \right\}$. In addition, during the each
integration, the degree increases by $1$ due to the fact that
$\int_0^{\mu_i} \mu_{i+1}^n d\mu_{i+1} = \mu_{i}^{n+1}/(n+1)$. Since
$\ell(1)-1$ variables from original $N-K$ variables of integration
vanished in $h\left( \mu_{\ell(1)}, \cdots, \mu_{\ell(K)} \right)$,
the degree to be added by the remaining variables of integration is
\begin{align}
D_{h,added} = N-K-\ell(1)+1. \label{eq:D_added}
\end{align}

The smallest degree of $r\left( \mu_{\ell(1)}, \cdots, \mu_{\ell(K)}
\right)$ is now ready to be calculated, which is
\ifCLASSOPTIONonecolumn
\begin{align}
D_{r,smallest} &= D_{g,smallest} + D_{h,smallest} \nonumber \\
&= D_{g,smallest} + \left(D_{h,org} - D_{h,vanishing} + D_{h,added} \right) \nonumber \\
&=(M-\ell(1)+1)(N-\ell(1)+1) - K \label{eq:D_smallest}
\end{align}
\else
\begin{align}
D_{r,smallest} &= D_{g,smallest} + D_{h,smallest} \nonumber \\
&= D_{g,smallest} \nonumber \\
& \quad + \left(D_{h,org} - D_{h,vanishing} + D_{h,added} \right) \nonumber \\
&=(M-\ell(1)+1)(N-\ell(1)+1) - K \label{eq:D_smallest}
\end{align}
\fi
where $D_{h,org} - D_{h,vanishing}$ stands for the degree of the
remaining non-vanishing variables of the term that leads to the
smallest degree of $h\left( \mu_{\ell(1)}, \cdots, \mu_{\ell(K)}
\right)$.

In the case of $\alpha_1 > 0$, the integrations over the variables
$\mu_i$, $1 \leq i \leq \ell(1)-1$ in (\ref{eq:function_h}) do not
exist. Contrary to the case of $\alpha_1 = 0$, no variable vanishes,
resulting in $D_{h,vanishing} = 0$, and $D_{h,added} = N-K$.
Equation (\ref{eq:D_smallest}) holds true for $\alpha_1 > 0$ since
$\ell(1) = 1$ in this case. Therefore, for any case of $\alpha_1$,
the smallest degree of the polynomial $r\left( \mu_{\ell(1)},
\cdots, \mu_{\ell(K)} \right)$ is $(M-\ell(1)+1)(N-\ell(1)+1)-K$.

\bibliographystyle{IEEEtran}
\bibliography{anal_BICMB.bbl}

\begin{thebibliography}{10}
\providecommand{\url}[1]{#1}
\csname url@samestyle\endcsname
\providecommand{\newblock}{\relax}
\providecommand{\bibinfo}[2]{#2}
\providecommand{\BIBentrySTDinterwordspacing}{\spaceskip=0pt\relax}
\providecommand{\BIBentryALTinterwordstretchfactor}{4}
\providecommand{\BIBentryALTinterwordspacing}{\spaceskip=\fontdimen2\font plus
\BIBentryALTinterwordstretchfactor\fontdimen3\font minus
  \fontdimen4\font\relax}
\providecommand{\BIBforeignlanguage}[2]{{%
\expandafter\ifx\csname l@#1\endcsname\relax
\typeout{** WARNING: IEEEtran.bst: No hyphenation pattern has been}%
\typeout{** loaded for the language `#1'. Using the pattern for}%
\typeout{** the default language instead.}%
\else
\language=\csname l@#1\endcsname
\fi
#2}}
\providecommand{\BIBdecl}{\relax}
\BIBdecl

\bibitem{jafarkhaniBook}
H.~Jafarkhani, \emph{Space-Time Coding: Theory and Practice}.\hskip 1em plus
  0.5em minus 0.4em\relax Cambridge University Press, 2005.

\bibitem{palomarTSP03}
D.~P. Palomar, J.~M. Cioffi, and M.~A. Lagunas, ``Joint tx-rx beamforming
  design for multicarrier {MIMO} channels: A unified framework for convex
  optimization,'' \emph{{IEEE} Trans. Signal Process.}, vol.~51, no.~9, pp.
  2381--2401, September 2003.

\bibitem{sengulTC06AnalSingleMultpleBeam}
E.~Sengul, E.~Akay, and E.~Ayanoglu, ``Diversity analysis of single and
  multiple beamforming,'' \emph{{IEEE} Trans. Commun.}, vol.~54, no.~6, pp.
  990--993, June 2006.

\bibitem{OrdonezTSP07}
L.~G. Ordonez, D.~P. Palomar, A.~Pages-Zamora, and J.~R. Fonollosa,
  ``High-{SNR} analytical performance of spatial multiplexing {MIMO} systems
  with {CSI},'' \emph{{IEEE} Trans. Signal Process.}, vol.~55, no.~11, pp.
  5447--5463, November 2007.

\bibitem{akayTC06BICMB}
E.~Akay, E.~Sengul, and E.~Ayanoglu, ``Bit interleaved coded multiple
  beamforming,'' \emph{{IEEE} Trans. Commun.}, vol.~55, no.~9, pp. 1802--1811,
  September 2007.

\bibitem{akayTC06BICMB_arxiv}
\BIBentryALTinterwordspacing
E.~Akay, H.~J. Park, and E.~Ayanoglu, ``On bit-interleaved coded multiple
  beamforming,'' arXiv: 0807.2464. [Online]. Available: \url{http://arxiv.org}
\BIBentrySTDinterwordspacing

\bibitem{proakis}
J.~G. Proakis, \emph{Digital Communications}, 4th~ed.\hskip 1em plus 0.5em
  minus 0.4em\relax McGraw-Hill, 2000.

\bibitem{haccounTCOM89}
D.~Haccoun and G.~Begin, ``High-rate punctured convolutional codes for
  {V}iterbi and sequential decoding,'' \emph{{IEEE} Trans. Commun.}, vol.~37,
  no.~11, pp. 1113--1125, November 1989.

\bibitem{SingletonTIT64}
R.~C. Singleton, ``Maximum distance q-nary codes,'' \emph{{IEEE} Trans. Inf.
  Theory}, vol.~10, no.~2, pp. 116--118, April 1964.

\bibitem{ErsinTCOMM08}
E.~Sengul, H.~J. Park, and E.~Ayanoglu, ``Bit-interleaved coded multiple
  beamforming with imperfect {CSIT},'' \emph{{IEEE} Trans. Commun.}, to appear.

\bibitem{conanTCOM84}
J.~Conan, ``The weight spectra of some short low-rate convolutional codes,''
  \emph{{IEEE} Trans. Commun.}, vol.~32, no.~9, pp. 1050--1053, September 1984.

\bibitem{beginTCOM90}
G.~Begin, D.~Haccoun, and C.~Paquin, ``Further results on high-rate punctured
  convolutional codes for viterbi and sequential decoding,'' \emph{{IEEE}
  Trans. Commun.}, vol.~38, no.~11, pp. 1922--1928, November 1990.

\bibitem{edelmanThesis}
A.~Edelman, ``Eigenvalues and condition numbers of random matrices,'' Ph.D.
  dissertation, MIT, Cambridge, MA, 1989.

\bibitem{Verdubook}
A.~M. Tulino and S.~Verd\'{u}, \emph{Random Matrix Theory and Wireless
  Communications}.\hskip 1em plus 0.5em minus 0.4em\relax Now Publishers, 2004.

\bibitem{KnoppTIT00}
R.~Knopp and P.~A. Humblet, ``On coding for block fading channels,''
  \emph{{IEEE} Trans. Inf. Theory}, vol.~46, no.~1, pp. 189--205, January 2000.

\bibitem{TarokhJSAIC99}
V.~Tarokh, H.~Jafarkhani, and A.~R. Calderbank, ``Space-time block coding for
  wireless communications: Performance results,'' \emph{{IEEE} J. Sel. Areas
  Commun.}, vol.~17, no.~3, pp. 451--460, March 1999.

\end{thebibliography}

\end{document}